\pdfoutput=1
\documentclass[oupdraft]{bio}
\usepackage{url}
\usepackage{amsmath}
\usepackage{amssymb}
\usepackage{mathtools}
\usepackage{bbm}
\usepackage{graphicx}%for plot extensions and plot inclusion
\usepackage{grffile} %for pdf extension with dots
\usepackage{setspace}
\usepackage[shortlabels]{enumitem}
\usepackage{geometry}
\usepackage{algorithm}
\usepackage{algorithmic}
\usepackage{booktabs}
\usepackage{afterpage}
\usepackage{float}
\usepackage{setspace}
\theoremstyle{prop}
\newtheorem{prop}{Proposition}
\begin{document}
\newcommand{\bSigma}{\boldsymbol\Sigma}
\newcommand{\btheta}{\boldsymbol\theta}
\newcommand{\Pow}{\mathrm{Power}_{\alpha,n}\left({\bSigma}, \boldsymbol\Delta, \Delta_{\mathrm{min}}\right)}
% Title of paper
\title{Power Analysis in a SMART Design: Sample Size Estimation for Determining the Best Dynamic Treatment Regime}
% List of authors, with corresponding author marked by asterisk
\author{ William J. Artman$^\ast$\\
\textit{Department of Biostatistics and Computational Biology, University of Rochester Medical Center, Rochester, NY 14620, USA}\\ \url{William_Artman@URMC.Rochester.edu} \\[0.2cm]Inbal Nahum-Shani\\\textit{Institute for Social Research, University of Michigan, Ann Arbor, MI 48106, USA}\\[0.2cm] Tianshuang Wu \\[0.2cm] James R. McKay \\\textit{
Department of Psychiatry, Perelman School of Medicine, University of Pennsylvania, Philadelphia, PA 19104, USA.} \\[0.2cm] Ashkan Ertefaie \\\textit{Department of Biostatistics and Computational Biology, University of Rochester Medical Center, Rochester, NY 14620, USA} 
}

% Running headers of paper:
\markboth%
% First field is the short list of authors
{Author and Others}
% Second field is the short title of the paper
{Power Analysis in a SMART design}

\maketitle
% Add a footnote for the corresponding author if one has been
% identified in the author list
%\footnotetext{To whom correspondence should be addressed.}
\vspace{1em}
\begin{abstract}
{Sequential, multiple assignment, randomized trial (SMART) designs have become increasingly popular in the field of precision medicine by providing a means for comparing sequences of treatments tailored to the individual patient, i.e., dynamic treatment regime (DTR). The construction of evidence-based DTRs promises a replacement to adhoc one-size-fits-all decisions pervasive in patient care. 
However, there are substantial statistical challenges in sizing SMART designs due to the complex correlation structure between the DTRs embedded in the design. Since the primary goal of SMARTs is the construction of an optimal DTR, investigators are interested in sizing SMARTs based on the ability to screen out DTRs inferior to the optimal DTR by a given amount which cannot be done using existing methods. In this paper, we fill this gap by developing a rigorous power analysis framework that leverages multiple comparisons with the best methodology. Our method employs Monte Carlo simulation in order to compute the minimum number of individuals to enroll in an arbitrary SMART. We will evaluate our method through extensive simulation studies. We will illustrate our method by retrospectively computing the power in the Extending Treatment Effectiveness of Naltrexone SMART study.
}
{DTR; Dynamic treatment regime; Monte Carlo; Multiple comparisons with the best; Power analysis; SMART design; }
\end{abstract}
\section{Introduction}
\label{sec1}

Sequential, multiple assignment, randomized trial (SMART) designs have gained considerable attention in the field of precision medicine by providing an empirically rigorous experimental approach for comparing sequences of treatments tailored to the individual patient, i.e., dynamic treatment regime (DTR) (\citealp{LAVORI2000605}; \citealp{murphy2005experimental}; \citealp{lei2012smart}). A DTR is a treatment algorithm implemented through a sequence of decision rules which dynamically adjusts treatments and dosages to a patient's unique changing need and circumstances (\citealp{murphy2001marginal}; \citealp{murphy2003optimal}; \citealp{robins2004optimal}; 
\citealp{nahum2012experimental}; \citealp{chakraborty2013statistical}; \citealp{chakraborty2014dynamic}; \citealp{laber2014dynamic}). SMART studies are motivated by scientific questions concerning the construction of an effective DTR. The sequential randomization in a SMART gives rise to several DTRs which are embedded in the SMART by design. Many SMARTs are designed to compare these embedded DTRs and identify those showing greatest potential for improving a primary clinical outcome. The construction of evidence-based DTRs promises an alternative to adhoc one-size-fits-all decisions pervasive in patient care. 

The advent of SMART designs poses interesting statistical challenges in the planning phase of the trials. In particular, performing power analyses to determine an appropriate sample size of individuals to enroll becomes mathematically difficult due to the complex correlation structure between the DTRs embedded in the design. Previous work in planning the sample size for SMARTs includes sizing pilot SMARTs (small scale versions of a SMART) so that each sequence of treatments has a pre-specified number of individuals with some probability by the end of the trial (\citealp{almirall2012designing}; \citealp{hwanwoo2016}). The central questions motivating this work are feasibility of the investigators to carry out the trial and acceptability of the treatments by patients. These methods do not provide a means to size SMARTs for comparing DTRs in terms of a primary clinical outcome. 

Alternatively, \citeauthor{CrivelloEvaluation2007} (2007(a)) proposed a new objective for SMART sample size planning. The question they address is how many individuals need to be enrolled so that the best DTR has the largest sample estimate with a given probability (\citealp{CrivelloStatMethod2007}(b)). Such an approach based on estimation alone fails to account for the uncertainty in the mean DTR sample estimates. In particular, some DTRs may be statistically indistinguishable from the true best for the given data in which case they should not necessarily be excluded as suboptimal. Furthermore, the true best DTR may only be superior in efficacy to other DTRs by an amount which is clinically insignificant at the cost of providing treatments that are more burdensome and costly compared to other DTRs. \cite{CrivelloEvaluation2007} also discussed sizing SMARTs to attain a specified power for testing hypotheses which compare only two treatments or two embedded DTRs as opposed to comparing all embedded DTRs. The work of \cite{CrivelloEvaluation2007} focused mainly on a particular common two-stage SMART design whereas our method is applicable to arbitrary SMART designs.

One of the main goals motivating SMARTs is to identify the optimal DTRs. It follows that investigators are interested in sizing SMARTs based on the ability to exclude DTRs which are inferior to the optimal DTR by a clinically meaningful amount. This cannot be done with existing methods. In this paper, we fill the current methodological gap by developing a rigorous power analysis framework that leverages the multiple comparisons with the best (MCB) methodology in order to construct a set of optimal DTR which excludes DTRs inferior to the best by a specified amount (\citealp{hsu1981simultaneous}; \citealp{hsu1984}; \citealp{hsu1996multiple}). Our method is applicable to an arbitrary SMART design. There are two main justifications for using multiple comparisons with the best. 1) MCB involves fewer comparisons compared to methods which involve all pairwise comparisons, and thus, it yields greater power for the same number of enrolled individuals with all else being equal (\citealp{ertefaie2015}); 2) MCB permits construction of a set of more than one optimal DTR from which the physician and patient may choose.

In Section 2, we give a brief overview of SMARTs and relevant background on estimation and MCB. In Section 3, we present our power analysis framework. In Section 4, we look at properties of our method. In Sections 5 and 6, we demonstrate the validity of our method through extensive simulation studies. In Section 7, we will apply our method to retrospectively compute the power in the Extending Treatment Effectiveness of Naltrexone SMART study. In the accompanying appendix, we provide additional details about estimation and our simulation study. In the future, an R package called ''smartsizer'' will be made freely available to assist investigators with sizing SMARTs.

\section{PRELIMINARIES}
\subsection{Sequential Multiple Assignment Randomized Trials}
\begin{figure}[h]
\centering
\includegraphics[width = 15cm]{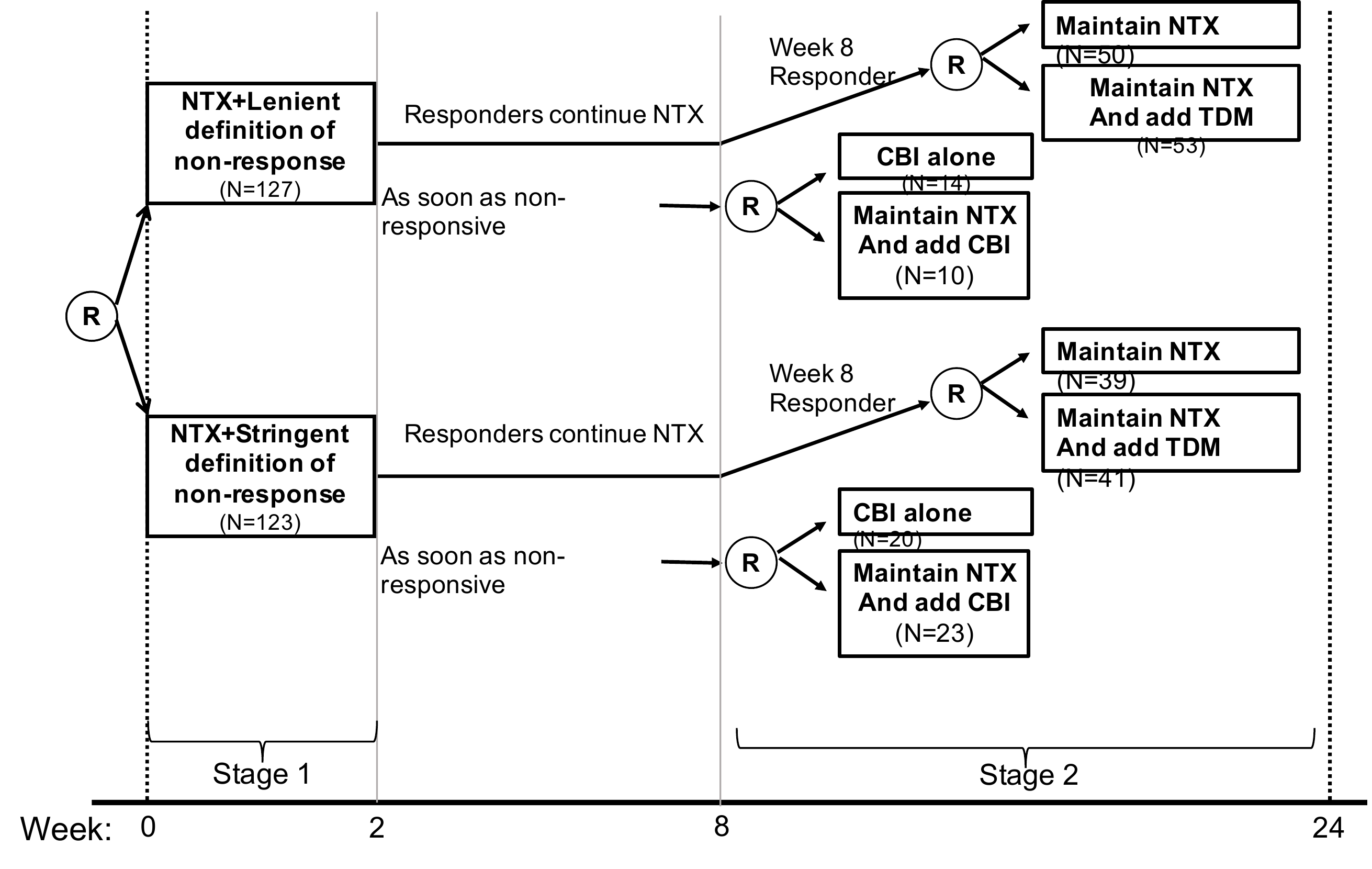}
\caption{This diagram show the structure of the EXTEND trial.}
\label{fig:EXTEND-figure}
\end{figure}

In SMART design, individuals proceed through multiple stages of randomization such that some or all individuals may be randomized more than once. Additionally, treatment assignment is often tailored to the individual's ongoing response status (\citealp{nahum2012experimental}). For example, in the Extending Treatment Effectiveness of Naltrexone (EXTEND) SMART study (see Figure \ref{fig:EXTEND-figure} for the study design and \citealp{nahum2017smart} for more details about this study), individuals were initially randomized to two different criteria of non-response: lenient or stringent. Specifically, all individuals received the same fixed dosage of naltrexone (NTX) -- a medication that blocks some of the pleasurable effects resulting from alcohol consumption. After the first two weeks, individuals were evaluated weekly to assess response status.

Individuals assigned to the lenient criterion were classified as non-responders as soon as they had five or more heavy drinking days during the first eight weeks of the study, whereas those assigned to the stringent criterion were classified as non-responders as soon as they had two or more heavy drinking days during the first eight weeks. As soon as participants were classified as non-responders, they transitioned to the second stage where they were randomized to two subsequent rescue tactics: either switch to combined behavioral intervention (CBI) or add CBI to NTX (NTX + CBI). At week 8, individuals who did not meet their assigned non-response criterion were classified as responders and re-randomized to two subsequent maintenance interventions: either add telephone disease management (TDM) to the NTX (NTX + TDM) or continue NTX alone. Note that the second stage treatment options in the SMART are tailored to the individual's response status. That is, individuals are randomized to different subsequent interventions depending on their early response status. This leads to a total of eight DTRs that are embedded in this SMART by design. For example, one of these DTRs recommends to start the treatment with NTX and monitor drinking behaviors weekly using the lenient criterion (i.e., 5 or more heavy drinking days) to classify the individual as a non-responder. As soon as the individual is classified as a non-responder, add CBI (NTX + CBI); if at week eight the individual is classified as a responder, add TDM (NTX + TDM). Many SMARTs are motivated by scientific questions pertaining to the comparison of the multiple DTRs that are embedded in the SMART by design.
For example, determining an optimal EDTR in the EXTEND may guide in evaluating a patient's initial response to NTX and in selecting the best subsequent treatment. 

In the next section, we discuss how the multiple comparison with the best (MCB) procedure (\cite{hsu1981simultaneous}, \cite{hsu1984}, \cite{hsu1996multiple}) can be used to address scientific questions concerning the optimal EDTR. 

\subsection{Determining a Set of Best DTRs using Multiple Comparison with the Best}
\label{MCB}
The MCB procedure can be used to identify a set of optimal EDTRs which cannot be statistically distinguished from the true best EDTR. 

In order to carry out MCB, we must be able to estimate the mean outcome of each EDTR.
 Let the true mean outcome vector of EDTRs be denoted by $\btheta=(\theta_1,...,\theta_N)^t$ where $N$ is the total number of DTRs embedded in the SMART design. Assume $\hat{\btheta}$ is a consistent estimator for $\btheta$ such that \begin{equation} \sqrt{n}(\hat{\btheta}-\btheta)\overset{d}{\rightarrow} N(\boldsymbol0,\bSigma), \label{eq:1} \end{equation} where $\bSigma$ is the asymptotic covariance matrix and $n$ is the total number of individuals enrolled in the SMART. Proposition 1 in Appendix A of the supplementary material presents two estimators for $\btheta$ which satisfy \eqref{eq:1}.
 
 The MCB procedure entails comparing the mean outcome of each EDTR to the mean outcome of the best EDTR standardized by the standard deviation of their difference. In particular, $\mathrm{EDTR}_i$ is considered statistically indistinguishable from the true best EDTR for the available data if and only if the standardized difference $\dfrac{\hat{\theta}_i-\hat{\theta}_j}{\sigma_{ij}}\geq -c_{i,\alpha}$ for all $j \neq i$ where $\sigma_{ij}=\sqrt{\text{Var}(\hat{\theta}_i-\hat{\theta}_j)}$ and $c_{i,\alpha}>0$ is chosen such that the set of best EDTRs includes the true best EDTR with at least a specified probability $1-\alpha$. Then, the set of best can be written as $\hat{\mathcal{B}}:=\{\mathrm{EDTR}_i:\hat{\theta}_i\geq \underset{j\neq i}{\max}[\hat{\theta}_j-c_{i,\alpha}\sigma_{ij}]\}$ where $c_{i,\alpha}$ depends on $\alpha$ and the covariance matrix $\bSigma$. 
The above $\alpha$ represents the type I error rate for excluding the best EDTR from $\hat{\mathcal{B}}$. To control the type I error rate, it suffices to consider the situation in which the true mean outcomes are all equal.

Then, a sufficient condition for the type I error rate to be at most $\alpha$ is to choose $c_{i,\alpha}$ so that the set of best DTRs includes each EDTR with probability at least $1-\alpha$:
\begin{equation}\mathbb{P}(\mathrm{EDTR}_i \in \hat{\mathcal{B}} \mid \theta_1=\cdots = \theta_N) =1-\alpha \text{ for all } i =1,...,N. \end{equation}
It can be shown it is sufficient for $c_{i,\alpha}$ to satisfy:
\begin{equation}\int \mathbb{P}\left(Z_j \leq z+c_{i,\alpha} \sigma_{ij} , \text{ for all } j =1,...,N\right)\mathrm{d}\phi(z)=1-\alpha, \label{eq:c}\end{equation}
where $\phi(z)$ is the marginal cdf of $Z_i$ and $(Z_1,...,Z_N)^t \sim N\left(\boldsymbol0, \bSigma \right)$. Observe that $c_{i,\alpha}>0$ is a function of $\bSigma$ and $\alpha\leq 0.5$, but not of the number of individuals $n$. The integral in \eqref{eq:c} is analytically intractable, but the $c_{i,\alpha}$ may be determined using Monte Carlo methods.

The MCB procedure has both applied and theoretical advantages which justify its use to compare DTRs that are embedded in a SMART. In particular, in the event that more than one EDTR is included in the set of best, MCB provides provides clinicians with a set of optimal treatment protocols to choose from in order to individualize a patient's treatment. Identifying a set of DTRs is useful in the situation where the treatment suggested by one DTR may not be feasible for the patient due to lack of insurance coverage, drug interactions, and/or allergies. In addition to the practical advantages, MCB provides a set with fewer DTRs compared to other methods since  fewer comparisons yields increased power to exclude inferior DTRs from the set of best. %Specifically, for a SMART design where N is the number of EDTRs, the MCB procedure involves only $N$ comparisons whereas, for example, pairwise multiple comparison procedures entail $\binom{N}{2}$ comparisons.
Another statistical advantage of MCB is that it incorporates the uncertainty in sample estimates by providing a set of optimal DTRs instead of simply choosing the largest sample estimate.

Although MCB has theoretical and applied merits over alternative approaches, it comes with the price of additional statistical challenges such as that imposed by the correlation structure between EDTRs expressed through $\bSigma$ and the numerical inconvenience of the $\max$ operator. The correlation structure between EDTR outcomes arises, in part, due to overlapping interventions in distinct DTRs and because patients' treatment histories may be consistent with more than a single DTR. For example, patients in distinct EDTRs of the EXTEND trial all receive naltrexone. Also, patients who are  are classified as responders in stage 2 and subsequently randomized to NTX will be consistent with two EDTRs: one where non-responders are offered CBI and one where non-responders are offered NTX + CBI. Hence, deriving a convenient univariate test statistic unfeasible and many computations analytically intractable. In order to overcome these challenges, we will derive a form of the power function for which Monte Carlo simulation can be applied in the following section. 

\section{Methods}
\label{method}
 
Let $\Delta_{\mathrm{min}}>0$ be the minimum detectable difference, $\alpha$ be the type I error rate, and $\bSigma$ be the asymptotic covariance matrix of $\sqrt{n}\hat{\btheta}$ where $n$ is the total number of individuals enrolled in the SMART. Furthermore, let $1-\beta$ denote the desired minimum power to exclude EDTRs with true outcome $\Delta_{\mathrm{min}}$ or more away from that of the true best outcome. Let $\boldsymbol \Delta$ be the vector of effect sizes. So, $\boldsymbol\Delta_i = \theta_N-\theta_i$. 

We wish to exclude all $i$ from the set of best for which $\boldsymbol\Delta_i\geq\Delta_{\mathrm{min}}$. We have that
\begin{align}
\mathrm{Power} &=\mathbb{P}_{\btheta,\bSigma, n}\left(\bigcap_{i:\boldsymbol\Delta_i\geq \Delta_{\mathrm{min}}} \left\{\hat{\theta}_i < \underset{j\neq i}{\max} [\hat{\theta}_j-c_{i,\alpha}\sigma_{ij}]\right\}\right).\label{eq:bound}
\end{align}

However, the $\max$ operator makes \eqref{eq:bound} analytically and computationally complicated, so we will instead bound the RHS of the following inequality:

\begin{equation}\mathbb{P}_{\bSigma, n}\left(\bigcap_{i:\boldsymbol\Delta_i\geq \Delta_{\mathrm{min}}}\left\{\hat{\theta}_i<\underset{j\neq i}{\max}[\hat{\theta}_j-c_{i,\alpha}\sigma_{ij}]\right\} \right)\geq\mathbb{P}_{\bSigma, n}\left(\bigcap_{i:\boldsymbol\Delta_i\geq \Delta_{\mathrm{min}}} \left\{\hat{\theta}_i < \hat{\theta}_N-c_{i,\alpha}\sigma_{iN} \right\}\right)\label{eq:4}. 
\end{equation}

This inequality follows from noting that for all $i$,
\begin{equation}
\left\{\hat{\theta}_i<\hat{\theta}_N-c_{i,\alpha}\sigma_{iN}\right\}\subset \left\{\hat{\theta}_i <\underset{j\neq i}{\max}[\hat{\theta}_j-c_{i,\alpha}\sigma_{ij}\right\}. \end{equation}
Theoretically, the bound obtained using \eqref{eq:4} may be conservative, but it is often beneficial to be conservative when conducting sample size calculations because of unpredictable circumstances such as loss to follow up, patient dropout, and/or highly skewed responses. 
Since the normal distribution is a location-scale family, the power only depends on the vector of effect sizes $\boldsymbol\Delta$ and not on $\btheta$. Henceforth, we will call the RHS of \eqref{eq:4} the power function and write $\Pow$. It follows that \begin{align}
\Pow &= \mathbb{P}_{\bSigma, n}\left(\bigcap_{i:\boldsymbol\Delta_i\geq \Delta_{\mathrm{min}}} \left\{\hat{\theta}_i < \hat{\theta}_N-c_{i,\alpha}\sigma_{iN} \right\} \right)\nonumber\\
&=\mathbb{P}_{\bSigma, n}\left(\bigcap_{i:\boldsymbol\Delta_i\geq \Delta_{\mathrm{min}}}\left\{\dfrac{\sqrt{n}(\hat{\theta}_i-(\hat{\theta}_N-\Delta_i))}{\sigma_{iN}\sqrt{n}} < -c_{i,\alpha} +\dfrac{\Delta_i\sqrt{n}}{\sigma_{iN}\sqrt{n}}\right\} \right) \nonumber\\
&=\mathbb{P}_{\bSigma, n}\left(\bigcap_{i:\boldsymbol\Delta_i\geq \Delta_{\mathrm{min}}}\left\{W_i< -c_{i,\alpha} +\dfrac{\Delta_i \sqrt{n}}{\sigma_{iN}\sqrt{n}}\right\}\right), \label{eq:7}
\end{align}

where $\boldsymbol W = (W_1,...,W_{M})^t \sim N\left(\boldsymbol0,\tilde{\bSigma}\right)$ and $\displaystyle \tilde{\bSigma}_{ij}=\text{Cov}\left(\dfrac{\sqrt{n}(\hat{\theta}_i-(\hat{\theta}_N-\Delta_i))}{\sigma_{iN}\sqrt{n}}, \dfrac{\sqrt{n}(\hat{\theta}_j-(\hat{\theta}_N-\Delta_j))}{\sigma_{jN}\sqrt{n}}\right)$, and $M$ is the number of indices $i:\Delta_i\geq\Delta_{\mathrm{min}}$.  
Note that $\displaystyle\boldsymbol W, c_{i,\alpha}, \text{ and } \sigma_{iN}\sqrt{n}=\sqrt{\Sigma_{ii}+\Sigma_{NN}-2\Sigma_{iN}}$ do not depend on $n$ since $\bSigma$ does not depend on $n$. If the standardized effect sizes $\delta_i$ are specified instead of $\Delta_i$, $\Delta_i$ may be computed as $\delta_i \sigma_{iN} \sqrt{n}$.

It follows that the power may be computed by simulating normal random variables and substituting the probability in \eqref{eq:7} with the empirical mean $\mathbb{P}_n$ of the indicator variable.  

Recall the main point of this paper is to assist investigators in choosing a sufficient number of individuals to enroll in a SMART. To this end, we will derive a method for inverting $\Pow$ with respect to $n$. In particular, we are interested in finding the minimum $n$ such that $\Pow \geq 1-\beta$.
We proceed by rewriting the power function expression:

\begin{align}
 \Pow & = \mathbb{P}_{\bSigma, n}\left(\bigcap_{i:\boldsymbol\Delta_i\geq \Delta_{\mathrm{min}}}  \left\{\hat{\theta}_i < \hat{\theta}_N-c_{i,\alpha}\sigma_{iN} \right\} \right)\nonumber\\
 &= \mathbb{P}_{\bSigma,n}\left(\bigcap_{i:\boldsymbol\Delta_i\geq \Delta_{\mathrm{min}}}  \left\{\dfrac{\sqrt{n}(\hat{\theta}_i-\hat{\theta}_N+\Delta_i+c_{i,\alpha}\sigma_{iN})}{\Delta_i}< \sqrt{n}\right\}\right)\nonumber\\
 &=\mathbb{P}_{\bSigma,n}\left(\bigcap_{i:\boldsymbol\Delta_i\geq \Delta_{\mathrm{min}}}  \left\{X_i< \sqrt{n}\right\}\right) \nonumber\\
 &=\mathbb{P}_{\bSigma,n}\left(\bigcap_{i:\boldsymbol\Delta_i\geq \Delta_{\mathrm{min}}} \left\{X_i< c^*_{1-\beta}\right\}\right),\label{eq:9}
\end{align}
where $\boldsymbol X = (X_1,...,X_{M})^t\displaystyle \sim N\left(\begin{pmatrix}c_{1,\alpha}\sigma_{1N}\sqrt{n}/\Delta_1\\c_{2,\alpha}\sigma_{2N}\sqrt{n}/\Delta_2\\\vdots\\c_{M,\alpha}\sigma_{MN}\sqrt{n}/\Delta_{M}\end{pmatrix},\tilde{\bSigma}\right)$, $\tilde{\bSigma}_{ij}=\mathrm{Cov}\left(\dfrac{\sqrt{n}\left(\hat{\theta}_i-\hat{\theta}_N\right)}{\Delta_i}, \dfrac{\sqrt{n}\left(\hat{\theta}_j-\hat{\theta}_N\right)}{\Delta_j}\right)$, $M$ is the number of indices $i:\Delta_i\geq\Delta_{\mathrm{min}}$, and $c^{*}_{1-\beta}$ is the $1-\beta$ equicoordinate quantile for the probability in \eqref{eq:9}. It follows from \eqref{eq:9} that $n = (c^{*}_{1-\beta})^2 \label{eq:10}$. Here, we write the quantile $c_{1-\beta}^*$ with an asterisk to distinguish it from the quantile $c_{i,\alpha}$ which controls the type I error rate. If the standardized effect sizes $\delta_i$ are specified instead of $\Delta_i$, $\Delta_i$ may be computed as $\delta_i \sigma_{iN} \sqrt{n}$.  
The constant $c^{*}_{1-\beta}$ may be computed using Monte Carlo simulation to find the inverse of equation \eqref{eq:9} after first computing the $c_{i,\alpha}$'s. The above procedure works because the $c_{i,\alpha}$'s do not change with $n$, so the distribution of $\boldsymbol X$ is constant as a function of $n$. Our approach for computing $n$ is an extension of the sample size computation method in the appendix of \cite{hsu1996multiple} to the SMART setting when $\bSigma$ is known. Algorithm 1 and Algorithm 2 use Monte Carlo simulation to calculate the power and sample size, respectively, as a function of the covariance matrix and effect sizes.
\begin{algorithm}[t]
  \caption{SMART Power Computation}\label{powercomp}
   \begin{algorithmic}[1]
\STATE Given $\bSigma =\mathrm{Var}(\sqrt{n}\hat{\btheta})$, compute $c_{i,\alpha}$ for $i=1,..,N$. 
\STATE Given $\boldsymbol \Delta$ and $\Delta_{\mathrm{min}}$, generate $\boldsymbol W^{(k)} =\left (W_1^{(k)},...,W_{M}^{(k)}\right)^t \sim N\left(\boldsymbol0,\tilde{\bSigma}\right),$ for $k =1,...,m$,  \\where $\displaystyle \tilde{\bSigma}_{ij}=\text{Cov}\left(\dfrac{\sqrt{n}(\hat{\theta}_i-(\hat{\theta}_N-\Delta_i))}{\sigma_{iN}\sqrt{n}}, \dfrac{\sqrt{n}(\hat{\theta}_j-(\hat{\theta}_N-\Delta_j))}{\sigma_{jN}\sqrt{n}}\right)$, $m$ is the number of Monte Carlo repetitions, and $M$ is the number of indices $i:\Delta_i\geq\Delta_{\mathrm{min}}$.
\STATE Compute the Monte Carlo probability \\$\displaystyle \mathrm{Power}_{MC,n,\alpha}\left(\bSigma,\boldsymbol\Delta,\Delta_{\mathrm{min}}\right) \approx \mathbb{P}_m \left[\mathbbm{1}\left(\bigcap_{i:\boldsymbol\Delta_i \geq \Delta}\left\{W_i < -c_{i,\alpha}+\dfrac{\Delta_i\sqrt{n}}{\sigma_{iN}\sqrt{n}}\right\}\right)\right] \text{ for some } m \in \mathbb{N}$\\$ \text{ where } \mathbb{P}_m \text{ denotes the empirical average.}$
\end{algorithmic}
\end{algorithm}

Algorithm 1 and Algorithm 2 will be implemented in an R package ''smartsizer'' freely available to download for assisting investigators in planning SMARTs. 

In the next section we will explore properties of the power function. We will see that the power is highly sensitive to the choice of the covariance matrix $\bSigma$. This underscores the importance of using prior information when conducting power calculations in order to obtain valid sample size and power predictions. Such information may be obtained from pilot SMARTs and physicians' knowledge of the variability in treatment response.

\section{Power Function Properties}
\label{properties}
Having derived an expression for the power function and provided an algorithm for its computation, we wish to explore some of its properties. In particular, it is important to examine how sensitive the power is to the choice of $\bSigma$ when $\bSigma$ is known. We will address the case in which $\bSigma$ is unknown in Section 5.
For simplicity, we consider the most conservative case in which the effect sizes are all equal: $\Delta_i =\Delta$ for all $i$. 

\begin{figure}[h]
\centering
\includegraphics[width = 10cm, trim = 0cm 3cm 0cm 0cm]{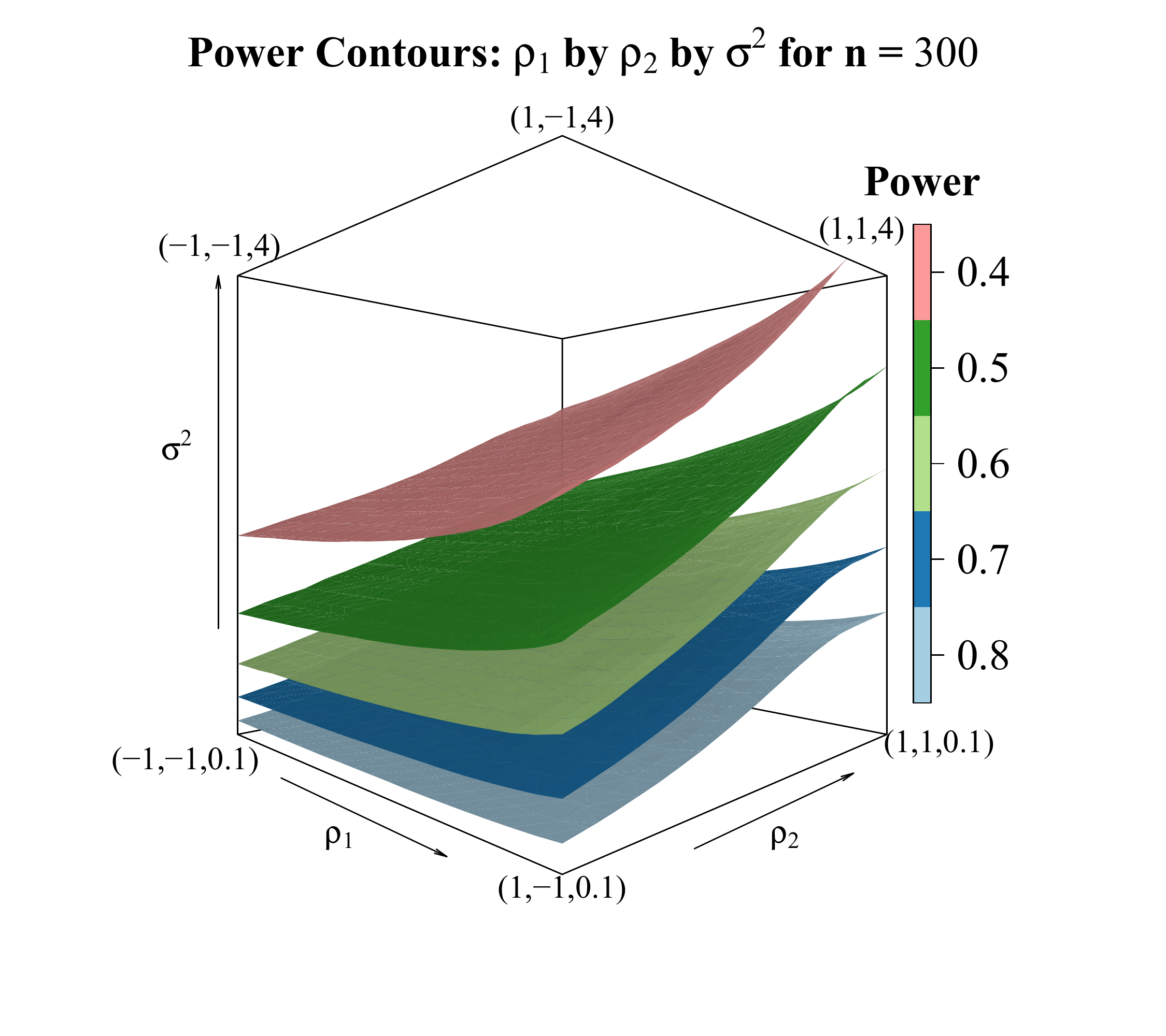}
\caption{
This plot shows the 3D contours of the power (denoted by shade/color) as a function of $\rho_1,\rho_2,\sigma^2$ where $\bSigma = \begin{pmatrix}1&\rho_1&0&0\\\rho_1&1&0&0\\0&0&1&\rho_2 \sigma\\0&0&\rho_2\sigma&\sigma^2\end{pmatrix}$ and $i=4$ is the best DTR. $\boldsymbol\Delta = (0.25, 0.25, 0.25, 0)$ and $\Delta_{\mathrm{min}} = 0.25$
}
\label{fig:3D-Plot}
\end{figure}

\begin{figure}[h]
\centering
\includegraphics[width = 8cm, trim = 0cm 1cm 0cm 0cm]{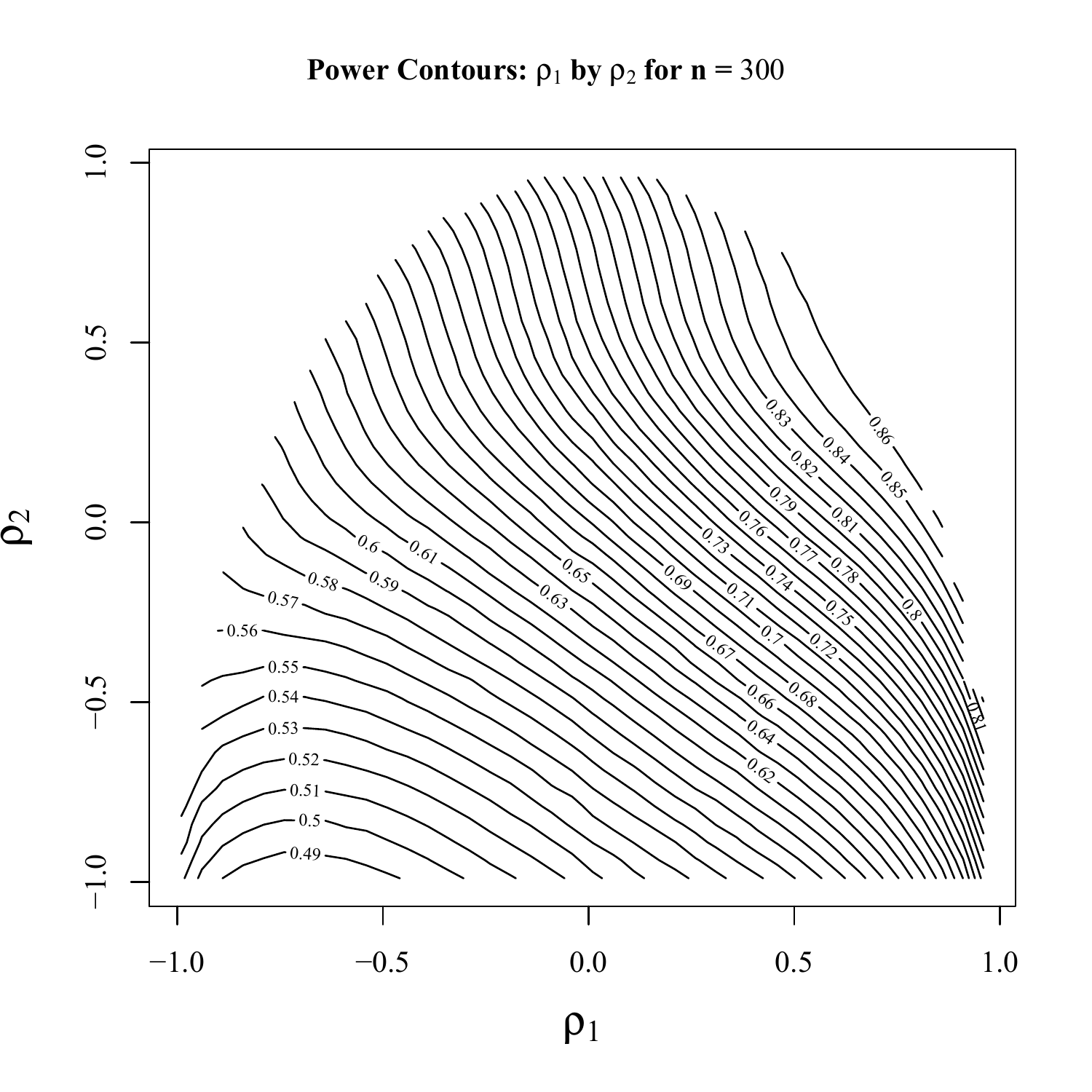}
\caption{$\bSigma =\begin{pmatrix} 1&0&0&0\\0&1&0&\rho_1\\0&0&1&\rho_2\\0&\rho_1&\rho_2&1\end{pmatrix}$ and the fourth EDTR is best. The effect sizes are set to $\Delta = 0.25$.
Note that the power appears monotone with respect to $\rho_1$ and $\rho_2$. The finger-shaped boundary is due to the feasible region of values for $\rho_1$ and $\rho_2$ such that $\bSigma$ is positive definite. 
}
\label{fig:best-non-best-contour}
\end{figure}

\begin{figure}[h]
\centering
\includegraphics[width = 9cm]{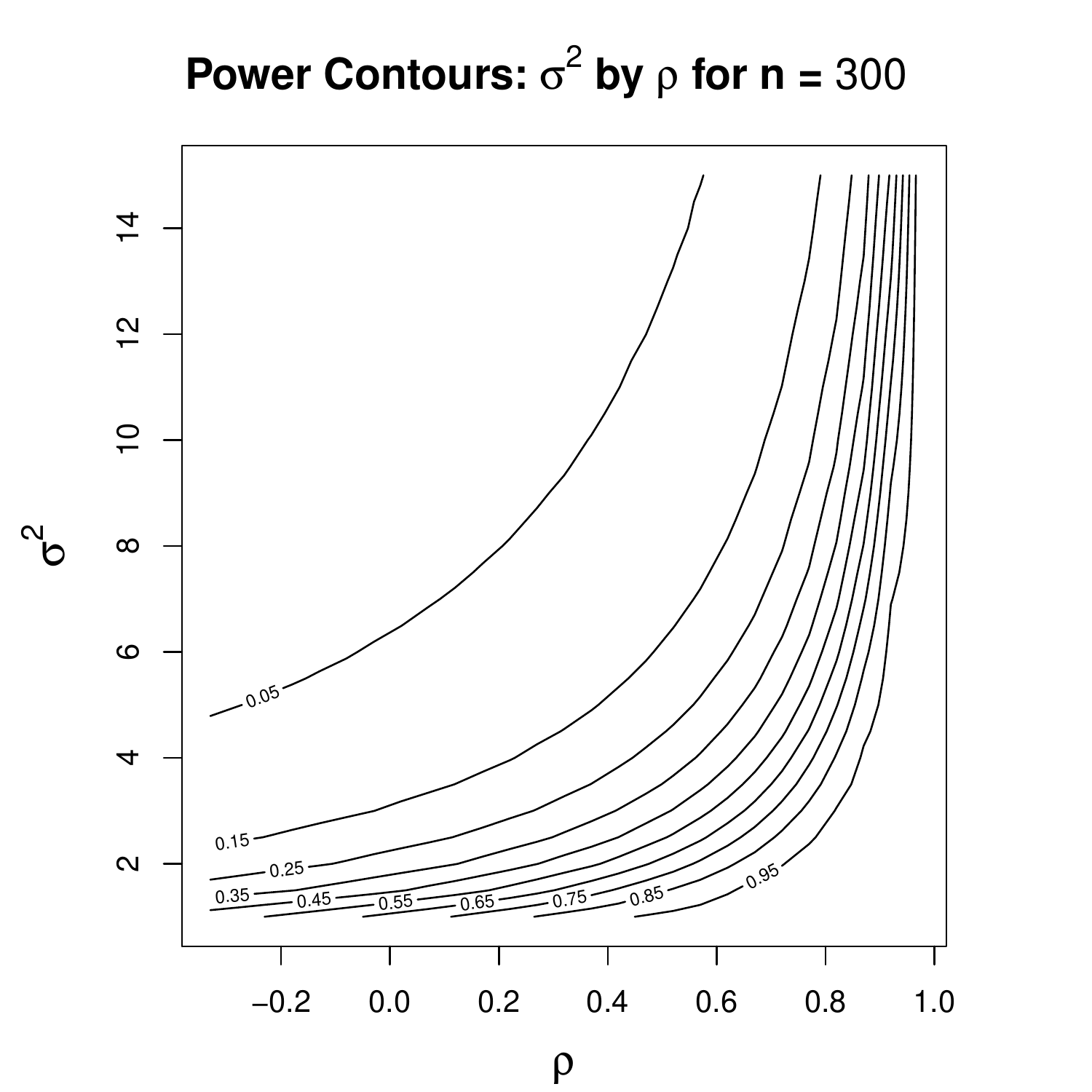}
\caption{Monotonicity in exchangeable $\bSigma$ case. $\bSigma = \begin{pmatrix}\sigma^2&\rho\sigma^2&\rho\sigma^2&\rho\sigma^2\\\rho\sigma^2&\sigma^2&\rho\sigma^2&\rho\sigma^2\\\rho\sigma^2&\rho\sigma^2&\sigma^2&\rho\sigma^2\\\rho\sigma^2&\rho\sigma^2&\rho\sigma^2&\sigma^2\end{pmatrix}$
}
\label{fig:contour-exchangeable}
\end{figure}

In Figures \ref{fig:3D-Plot}-\ref{fig:contour-exchangeable}, we evaluated the power function over a grid of values for $\bSigma$ using Equation 3.6 and Algorithm 1. These plots suggest the trend that higher correlations and lower variances tend to yield higher power for both the exchangeable and non-exchangeable $\bSigma$. The correlation between best and non-best DTRs appears to have a greater influence on power than the correlation between two inferior DTRs as we see in Figure \ref{fig:3D-Plot}. 

\begin{algorithm}[t]
  \caption{SMART Sample Size Estimation}\label{samplesize}
   \begin{algorithmic}[1]
\STATE Given $\bSigma =\mathrm{Var}(\sqrt{n}\hat{\btheta})$, compute $c_{i,\alpha}$ for $i=1,..,N$.
\STATE Given $\boldsymbol\Delta$ and $\Delta_{\mathrm{min}}$, generate $\boldsymbol X^{(k)} = \left(X_1^{(k)},...,X_{M}^{(k)}\right)^t\displaystyle \sim N\left(\begin{pmatrix}c_{1,\alpha}\sigma_{1N}\sqrt{n}/\Delta_1\\c_{2,\alpha}\sigma_{2N}\sqrt{n}/\Delta_2\\\vdots\\c_{M,\alpha}\sigma_{MN}\sqrt{n}/\Delta_{M}\end{pmatrix},\tilde{\bSigma}\right),$ \\for $k =1,...,m,$ where $\tilde{\bSigma}_{ij}=\mathrm{Cov}\left(\dfrac{\sqrt{n}\left(\hat{\theta}_i-\hat{\theta}_N\right)}{\Delta_i}, \dfrac{\sqrt{n}\left(\hat{\theta}_j-\hat{\theta}_N\right)}{\Delta_j}\right)$, $m$ is the number of Monte Carlo repetitions, and $M$ is the number of indices $i:\Delta_i\geq\Delta_{\mathrm{min}}$.
\STATE Find the empirical $1-\beta$ equicoordinate quantile $c^{*}_{1-\beta}$ of the simulated $\boldsymbol X^{(k)}$ for each $k=1,...,m$.
\STATE Then, the sample size is $n \approx \left(c^*_{1-\beta}\right)^2$
\end{algorithmic}
\end{algorithm}

It is analytically difficult to prove monotonicity for a general $\bSigma$ structure. However, as Figure \ref{fig:contour-exchangeable} suggests, it can be proven the power function is a monotone non-decreasing function of the correlation and a monotone non-increasing function of the variance for an exchangeable covariance matrix. We conjecture this property is true in general for $n$ sufficiently large. We state the result below.

\begin{theorem}
Let $\bSigma$ be exchangeable: $\bSigma = \sigma^2\boldsymbol I_{N} + \rho\sigma^2 \left( \mathbbm{1}_N \mathbbm{1}_{N}'-\boldsymbol I_N \right)$ where $\boldsymbol I_N =\mathrm{diag}(1,...,1)$ and $\mathbbm{1}_N = (1,...,1)'$. 
Then, the power is a monotone increasing function of $\rho$ and a monotone decreasing function of $\sigma^2$.
\end{theorem}

In order to simplify computations, we propose choosing a covariance matrix with fewer parameters than the posited covariance matrix. That is, suppose $E$ denotes the space of  covariance matrices with fewer parameters, and that the posited covariance matrix for the SMART is $\bSigma_{\mathrm{True}}$. For example, $E$ may be the space of exchangeable covariance matrices. We suggest choosing $\bSigma_{\mathrm{Exchangeable}} = \underset{\bSigma\in E}{\arg\min} \left\lVert \bSigma_{\mathrm{True}}- \bSigma\right\rVert$. For example, $\lVert\cdot\rVert$ may be the Frobenius norm. In our simulation study, we will evaluate the power when choosing the nearest exchangeable matrix in the Frobenius norm.

\begin{figure}[h]
\centering
\includegraphics[width = 15cm]{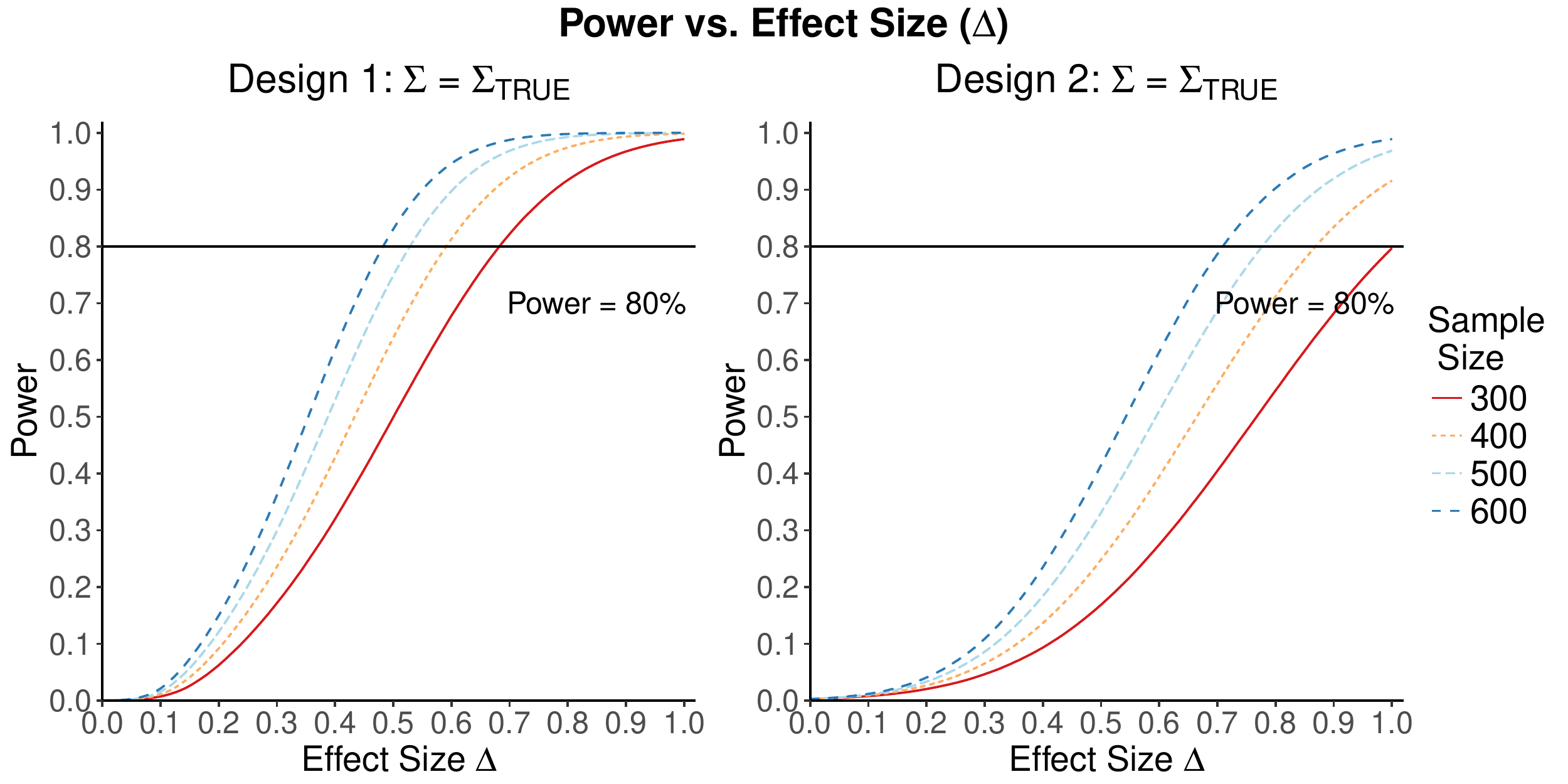}
\caption{The plot of power against the uniform effect size for $\boldsymbol\Sigma = \boldsymbol I_4,\bSigma_{\mathrm{TRUE}}$ shows that as the effect size $\Delta \to \infty$, $\mathrm{Power}(\Delta) \to 1$ and similarly $\mathrm{Power}(\Delta) \to 0$ as $\Delta \to 0$. $\bSigma_{\mathrm{TRUE}}$ is the true covariance matrix form simulation design 1. Furthermore, the plot demonstrates the power is a monotone increasing function of the effect size $\Delta$. A proof can be derived using continuity and monotonicity of the probability measure.}
\label{fig:power-by-delta}
\end{figure}

We also investigated the power as a function of the effect size assuming all effect sizes are equal by computing over a grid of $\Delta$ values. Figure \ref{fig:power-by-delta} shows how the power is a monotone increasing function of the uniform effect size $\Delta$ which makes intuitive sense.

\section{Simulation Study}
We have explored how the power changes in terms of a known covariance matrix and effect size. In this section, we present simulation studies for two different SMART designs in which we evaluate the assumption of a known covariance matrix. In practice, the true covariance matrix is estimated consistently by some $\hat{\bSigma}$. The designs are based on those discussed in \cite{ertefaie2015}. For each SMART, we simulate 1000 datasets across a grid of sample sizes $n$.
We computed the sets of best EDTRs using the estimates $\hat{\btheta}$ and $\hat{\bSigma}$ obtained from the AIPW estimation method after correctly specifying an appropriate marginal structural model and conditional means (see Section 1.2, Proposition 1 of Appendix A in the supplementary material, and \cite{ertefaie2015} for more details). For each $n$, the empirical power was calculated as the fraction of data sets which excluded all EDTRs with true mean outcome $\Delta_{\mathrm{min}}$ or more away from the best EDTR.

\subsection{SMART Design: Example 1}
In design 1, the stage-2 randomization is tailored based on response to the stage-1 treatment assignment. The tailoring variable is the indicator $V \in \{R,NR\}$ where a response corresponds to the intermediate outcome $O_2$ being positive. Non-responders to the first stage treatment are subsequently re-randomized to one of two intervention options while responders continue on the initial treatment assignment. See Figure \ref{fig:SMART-design-1} in Appendix B of the supplementary material for more details. We generated 1000 data sets for each sample size $n = 50,100,150,200,250,300,350,400,450,500$ according to the following scheme:
\begin{enumerate}
\itemsep0em 
\item \begin{enumerate}
\itemsep0em 

\item Generate $O_{11}, O_{12} \sim N(0,1)$ (baseline covariates)
\item Generate $A_1 \in \{-1,+1\}$ from a Bernoulli distribution with probability $0.5$ (first-stage treatment option indicator)
\end{enumerate}
\item \begin{enumerate}
\itemsep0em 
\item Generate $O_{21} \mid O_{11} \sim N(0.5 O_{11},1)$ and $O_{22} \mid O_{12} \sim N(0.5O_{12}, 1)$ (intermediate outcomes) 
\item Generate $A_2^{NR} \in \{-1,+1\}$ from a Bernoulli distribution with probability $0.5$ (second-stage treatment option indicator for non-responders)
\end{enumerate}
\item \small $Y \mid O_{11},O_{12},O_{21},O_{22},A_1,A_2^{NR} \sim N\left(1+O_{11}-O_{12}+O_{22}+O_{21}+A_1(\delta+\frac{O_{11}}{2})+I(O_{21}>0)A_2^{NR}\dfrac{\delta}{2},1\right)$\\where $\delta=0.25$
\end{enumerate}
The parameter estimates $\hat{\beta}_{\mathrm{AIPW}}$ were computed using augmented-inverse probability weighting (AIPW) (see Appendices A and B in the supplementary material and \citealp{ertefaie2015} for more details).

Then, $\hat{\btheta}_{\mathrm{AIPW}} = \boldsymbol D \hat{\boldsymbol\beta}_{\mathrm{AIPW}}$ where 
\begin{equation}
\boldsymbol D = \begin{pmatrix}1&1&1\\1&-1&1\\1&1&-1\\1&-1&-1\end{pmatrix}.
\end{equation} The rows of $\boldsymbol D$ correspond to each of the four EDTRs as listed in the supplementary material.
 
The true $\btheta$ was $(1.312,0.812,1.188,0.688)$. The minimum effect size $\Delta_{\mathrm{min}}$ was set to $0.5$ and the vector of effect sizes was set to $(0, 0.5, 0.124, 0.624)$ for estimating the anticipated power. We computed the set of best DTRs using the multiple comparison with the best procedure as outlined in Section 2.4 for each data set and sample size. The empirical power was calculated as the fraction of data sets which excluded all EDTRs with true mean outcome $\Delta_{min}$ or more away from the best EDTR (in this case $\mathrm{EDTR}_2$ and $\mathrm{EDTR}_4$), for each $n$.
The true covariance matrix for this SMART was estimated by averaging 1000 simulated datasets each of 10000 individuals. $\bSigma_{\mathrm{Exchangeable}}$ is the closest matrix in Frobenius norm to $\bSigma_{\mathrm{True}}$ of the form 
\begin{equation*}
\begin{pmatrix} \sigma^2 &\rho\sigma^2&\rho\sigma^2&\rho\sigma^2\\\rho\sigma^2&\sigma^2&\rho\sigma^2&\rho\sigma^2\\\rho\sigma^2&\rho\sigma^2&\sigma^2&\rho\sigma^2\\\rho\sigma^2&\rho\sigma^2&\rho\sigma^2&\sigma^2\end{pmatrix}
\end{equation*}

\subsubsection{Example 1: results}
\label{sim1results}

\begin{figure}[h]
\centering
\includegraphics[width = 8cm]{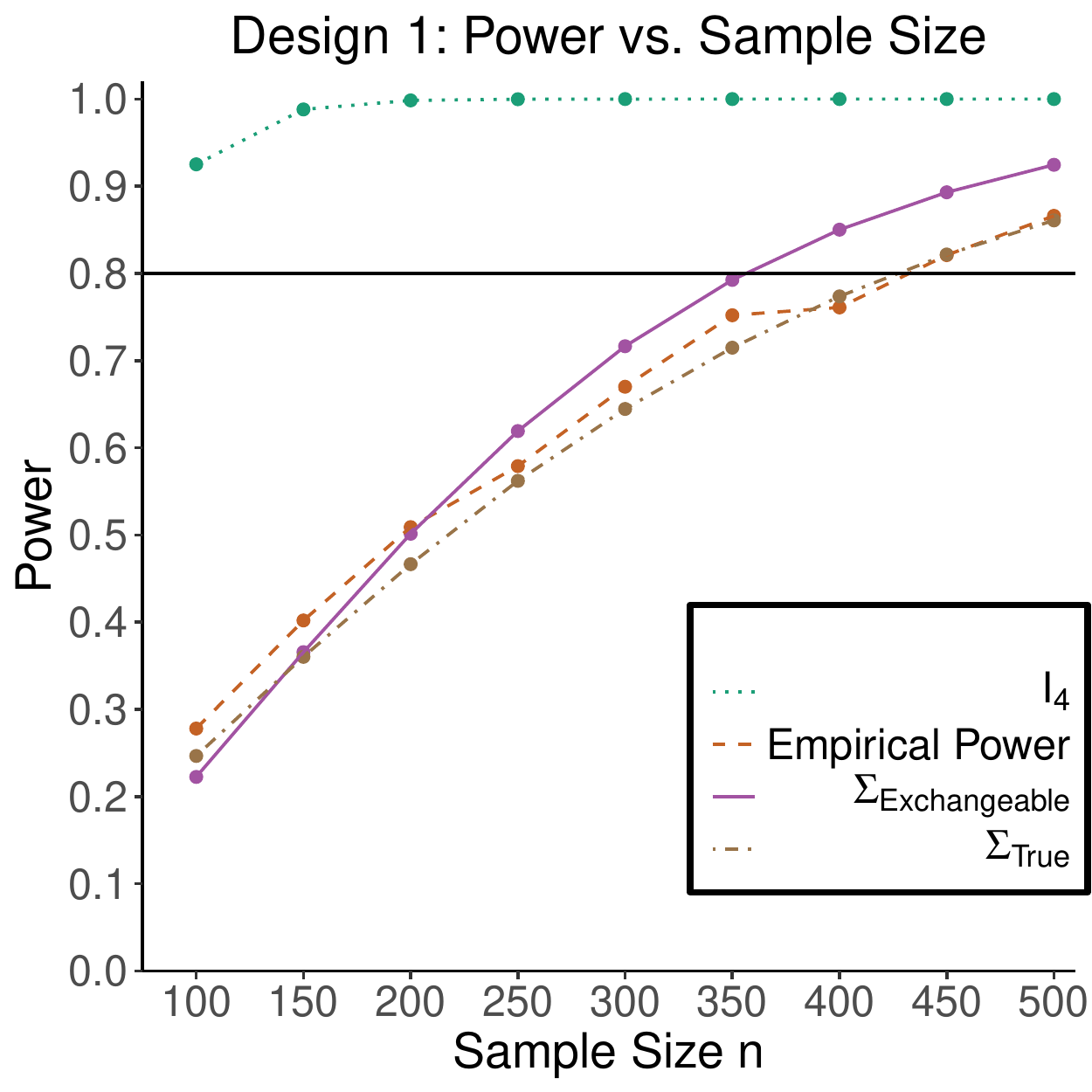}
\includegraphics[width = 8cm]{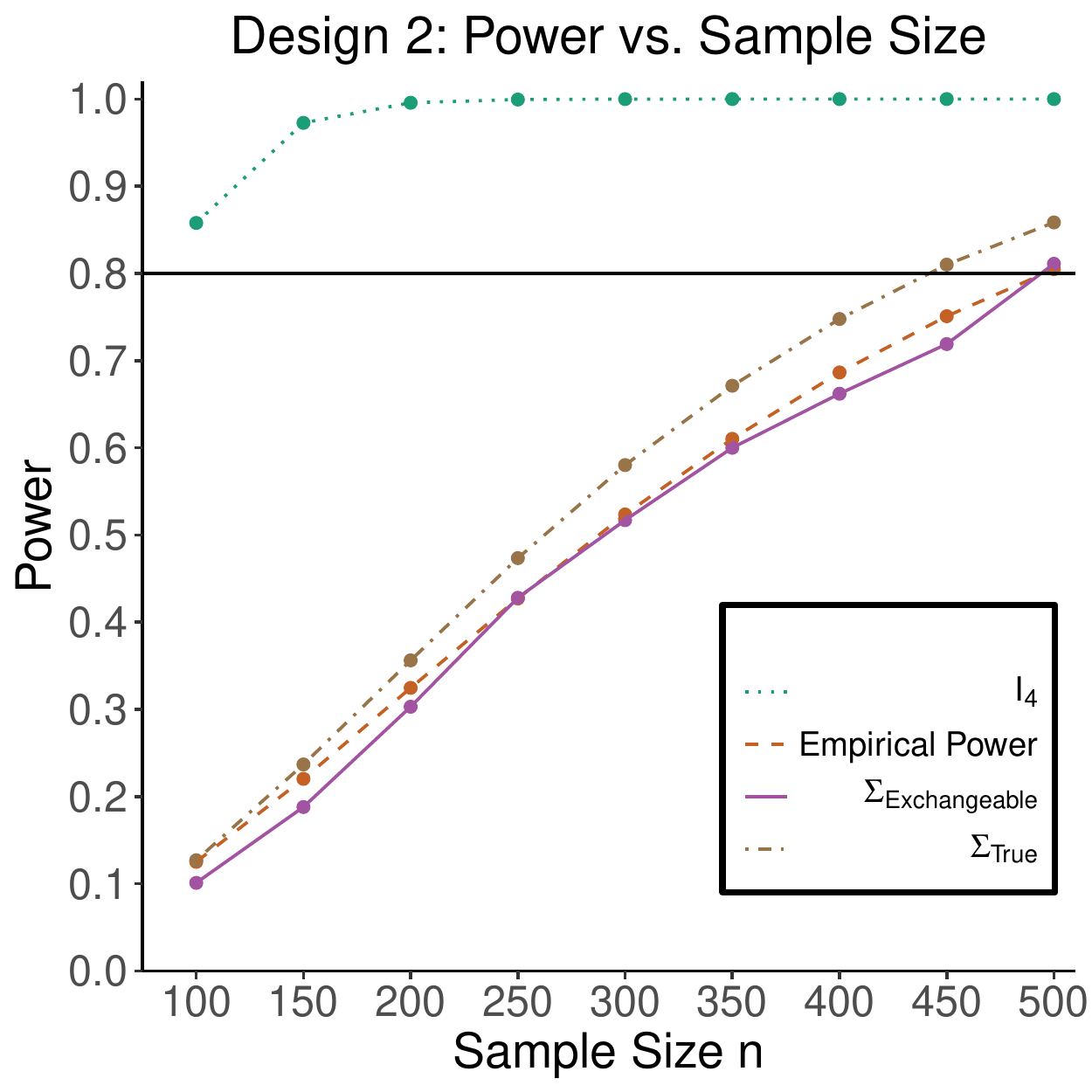}
\caption{The plots shows the power as a function of the sample size n with a horizontal line where the power is $80\%$. The plot shows the power curves for $\bSigma = \boldsymbol I_4, \bSigma=\bSigma_{\mathrm{True}}$, 
 and $\bSigma = \bSigma_{\mathrm{Exchangeable}}$ and the empirical power curve.}
\label{fig:power-plot}
\end{figure}

The simulation results are summarized in the plot on the left-hand side of Figure \ref{fig:power-plot}. The plot shows the sample size calculation is highly sensitive to the choice of $\bSigma$. Choosing $\bSigma = \boldsymbol I_4$ will greatly underestimate the required sample size, predicting around 100 individuals compared to the true 400-450 individuals needed to achieve $80\%$ power. This implies it is important not to ignore the variance of DTRs. 
Consequently, it is very important to utilize prior information such as that obtained from pilot SMARTs and physician's knowledge of treatment response variability in order to make an informed choice for the covariance matrix and to obtain accurate sample size predictions.

\subsection{SMART Design: Example 2}
In the second SMART design, stage-2 randomization depends on both prior treatment and intermediate outcomes (\citealp{ertefaie2015}). In particular, Individuals are randomized at stage-2 if and only if they are non-responders whose stage-1 treatment option corresponded to $A_1 =-1$ (call this condition B) (see Appendix B Figure \ref{fig:SMART-design-2} for more details).
 We generated 1000 data sets for each sample size $100,150,200,250,300,350,400,450,500$ according to the following scheme:
\begin{enumerate}
\itemsep0em 
\item \begin{enumerate}
\itemsep0em 

\item Generate $O_{11}, O_{12} \sim N(0,1)$ (baseline covariates)
\item Generate $A_1 \in \{-1,+1\}$ from a Bernoulli distribution with probability $0.5$ (first-stage treatment option indicator)
\end{enumerate}
\item \begin{enumerate}
\itemsep0em 
\item Generate $O_{21} \mid O_{11} \sim N(0.5 O_{11},1)$ and $O_{22} \mid O_{12} \sim N(0.5O_{12}, 1)$ (intermediate outcomes) 
\item Generate $A_2^{B} \in \{1,2,3,4\}$ from a Multinomial distribution with probability $0.25$ (second-stage treatment option indicator for individuals satisfying condition B)
\end{enumerate}
\item $Y \mid O_{11},O_{12},O_{21},O_{22},A_1,A_2^{B} \sim$ Normal with unit variance and mean equal to \\\small \begin{equation*}\begin{split}1&+O_{11}-O_{12}+O_{21}+O_{22}+I(A_1=-1)(\delta+O_{11})\\&+I(O_{21}>0)I(A_1=-1)[-\delta/4I(A_2=1)+\delta/2I(A_2=2)+0I(A_2=3)+\delta/2O_{21}I(A_2=2)]\end{split}\end{equation*}\\where $\delta=2.00$
\end{enumerate}
 The parameter estimates $\hat{\beta}_{\mathrm{AIPW}}$ were computed using AIPW (see Appendices A and B in the supplementary material and \citealp{ertefaie2015} for more details).
Then, $\hat{\btheta}_{\mathrm{AIPW}} = \boldsymbol D \hat{\boldsymbol\beta}_{\mathrm{AIPW}}$ where 
\begin{equation}\boldsymbol D =\begin{pmatrix}1&0&0&0&0\\1&1&0&0&0\\1&1&1&0&0\\1&1&0&1&0\\1&1&0&0&1\end{pmatrix}.
\end{equation}
 The true $\boldsymbol\beta$ value is $(1.00,\delta,-\delta/8,\delta/4,0)$ where $\delta = 2.00$ and the true $\btheta$ value is $(1.00,3.00,2.75,3.50,3.00)$.
 Note the fourth EDTR is the best. 
 
We let the vector of effect sizes $\Delta = (2.50,0.50, 0.75, 0.00, 0.50)$ and the desired detectable effect size $\Delta_{\mathrm{min}}=0.5$. The set of best was computed for each data set. For each sample size, the empirical power is the fraction of $1000$ data sets which exclude $\mathrm{EDTR}_1, \mathrm{EDTR}_2,\mathrm{EDTR}_3,\mathrm{EDTR}_5$.\\
 
\subsubsection{Example 2: Results}
 Our simulation results are summarized in the plot on the right-hand size of Figure \ref{fig:power-plot}. The true covariance matrix for this SMART was computed by averaging 1000 simulated datasets each of 10000 individuals. The power plots show that the predicted power is similar to the empirical power when assuming both the correct $\bSigma_{\mathrm{True}}$ and for $\bSigma$ of the form \begin{equation*}\begin{pmatrix}\sigma_1^2 &\rho_1\sigma_1\sigma_2&\rho_1\sigma_1\sigma_2&\rho_1\sigma_1\sigma_2&\rho_1\sigma_1\sigma_2\\
 \rho_1\sigma_1\sigma_2 &\sigma_2^2&\rho_2\sigma_2^2&\rho_2\sigma_2^2&\rho_2\sigma_2^2\\
 \rho_1\sigma_1\sigma_2&\rho_2\sigma_2^2&\sigma_2^2&\rho_2\sigma_2^2&\rho_2\sigma_2^2\\
 \rho_1\sigma_1\sigma_2&\rho_2\sigma_2^2&\rho_2\sigma_2^2&\sigma_2^2&\rho_2\sigma_2^2\\
 \rho_1\sigma_1\sigma_2&\rho_2\sigma_2^2&\rho_2\sigma_2^2&\rho_2\sigma_2^2&\sigma_2^2
 \end{pmatrix}
 \end{equation*}
 with parameters $\rho_1,\rho_2,\sigma_1^2,\sigma_2^2$ chosen to minimize the Frobenius distance to $\bSigma_{\mathrm{True}}$ (see Appendix B of the supplementary materials for more details). The anticipated sample size is approximately 500 individuals for $\bSigma_{\mathrm{True}}$.

\section{Illustration: EXTEND Retrospective Power Computation}

In this section, we will apply our power analysis method to examine how much power there was to distinguish between DTRs $\Delta_{\mathrm{min}}$ away from the best in the EXTEND SMART when using MCB. Please see Section 2.1 for more details about the EXTEND SMART and Figure \ref{fig:EXTEND-figure} for a diagram depicting the EXTEND SMART. The outcome of interest was the Penn Alcohol Craving Scale (PACS) and the lower PACS were considered better outcomes. The covariance matrix $\hat{\bSigma}$ and the vector of EDTR outcomes $\btheta$ were estimated using both IPW and AIPW (see Appendix A and Proposition 1 of the supplementary material for more details about estimation procedures). The covariance matrices are given below: 

\begin{align*}
\hat{\bSigma}_{\mathrm{IPW}} = \mathrm{Var}(\sqrt{n}\hat{\btheta}_{\mathrm{IPW}}) &= \begin{pmatrix} 145.86&54.88&77.24&-13.74&101.55&10.57&32.93&-58.05\\
54.88&163.02&1.13&109.27&12.66&120.8&-41.09&67.05\\
77.24&1.13&125.54&49.42&33.34&-42.77&81.64&5.53\\
-13.74&109.27&49.42&172.43&-55.55&67.46&7.62&130.63\\
101.55&12.66&33.34&-55.55&138.36&49.47&70.16&-18.73\\
10.57&120.8&-42.77&67.46&49.47&159.71&-3.87&106.37\\
32.93&-41.09&81.64&7.62&70.16&-3.87&118.86&44.84\\
-58.05&67.05&5.53&130.63&-18.73&106.37&44.84&169.94
\end{pmatrix}\\
\hat{\bSigma}_{\mathrm{AIPW}} = \mathrm{Var}(\sqrt{n}\hat{\btheta}_{\mathrm{AIPW}})&=\begin{pmatrix} 113.35&32.52&82.01&1.19&103.8&22.97&72.46&-8.36\\
32.52&143.74&-13.93&97.28&25.91&137.12&-20.55&90.67\\
82.01&-13.93&123.63&27.69&72.32&-23.63&113.94&17.99\\
1.19&97.28&27.69&123.78&-5.58&90.52&20.92&117.02\\
103.8&25.91&72.32&-5.58&112.1&34.21&80.62&2.73\\
22.97&137.12&-23.63&90.52&34.21&148.36&-12.39&101.76\\
72.46&-20.55&113.94&20.92&80.62&-12.39&122.09&29.08\\
-8.36&90.67&17.99&117.02&2.73&101.76&29.08&128.11
\end{pmatrix}
\end{align*}
The EDTR outcome vectors $\hat{\btheta}_{\mathrm{IPW}}$ and $\hat{\btheta}_{\mathrm{AIPW}}$ are summarized in Table 1.

The set of best when performing estimation using AIPW excluded $\mathrm{EDTR}_6$ and $\mathrm{EDTR}_8$, both of which had effect size greater than $2$ away from the best EDTR, $\mathrm{EDTR}_1$. The set of best when using IPW failed to exclude any of the inferior EDTR (\citealp{ertefaie2015}). In order to evaluate the power there was to exclude $\mathrm{EDTR}_6$ and $\mathrm{EDTR}_8$ in EXTEND when using AIPW, we set the minimum detectable effect size $\Delta_{\mathrm{min}}$ to $2$.

At an $\alpha$ level of $0.05$, the power to exclude all DTRs inferior to the best one by $2$ or more was $27\%$ for IPW and $46\%$ for AIPW. AIPW yields greater power than IPW because AIPW yields smaller standard errors compared with IPW (\citealp{ertefaie2015}). Our method estimates that a total of $717$ individuals would need to be enrolled to achieve a power of $80\%$ using IPW and a total of $482$ individuals would need to be enrolled when using AIPW. 

\begin{figure}[t]
\centering
\includegraphics[width = 10cm]{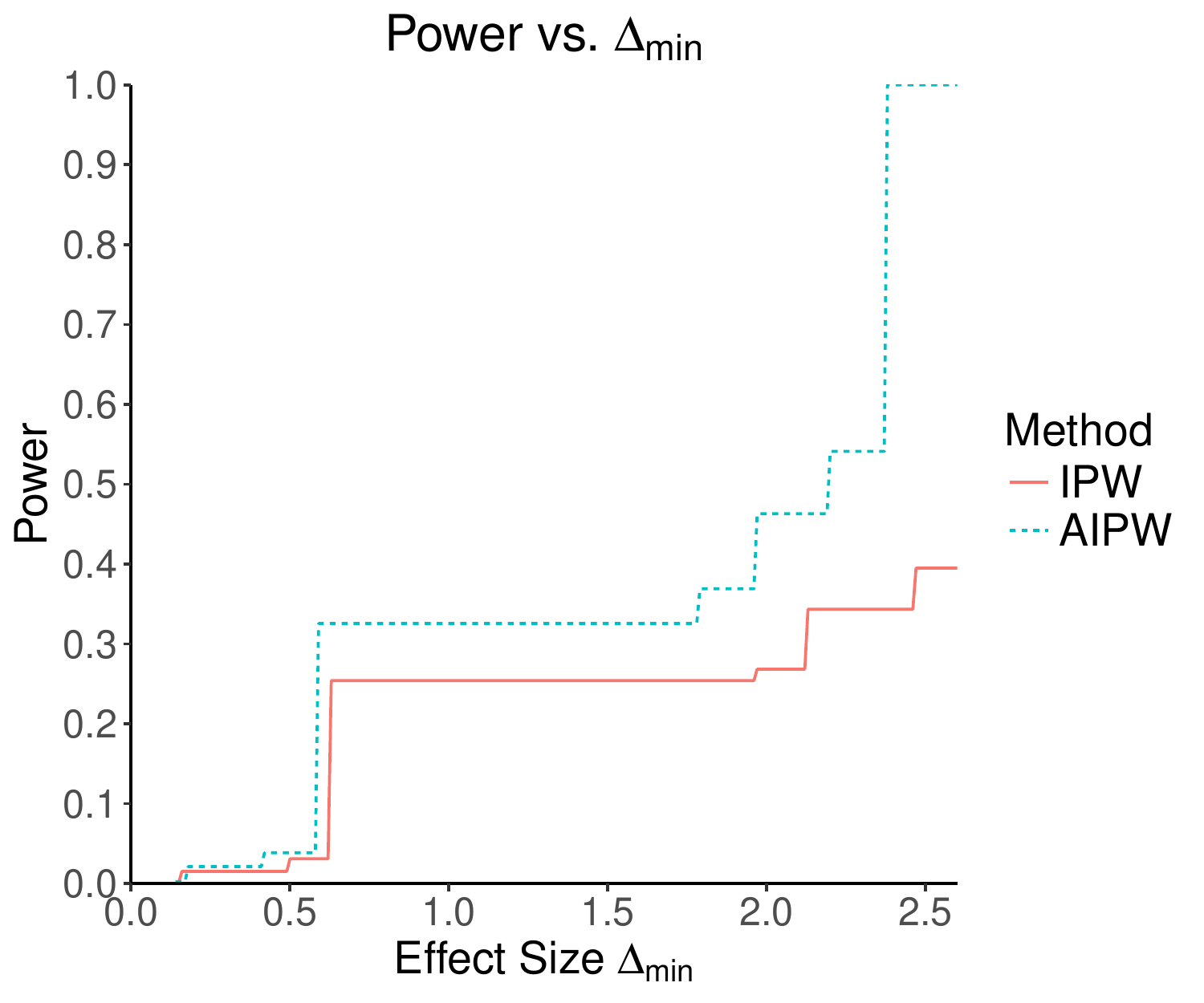}
\caption{This plot shows the power as a function of $\Delta_{\mathrm{min}}$ in the EXTEND trial when performing estimation with IPW and AIPW, respectively. There are 250 individuals in EXTEND}
\label{fig:EXTEND-power-step}
\end{figure}

In Figure \ref{fig:EXTEND-power-step}, we computed the power over a grid of $\Delta$ values to see how the power changes as a function of effect size. 

\begin{figure}[H]
\centering
\includegraphics[width = 10cm]{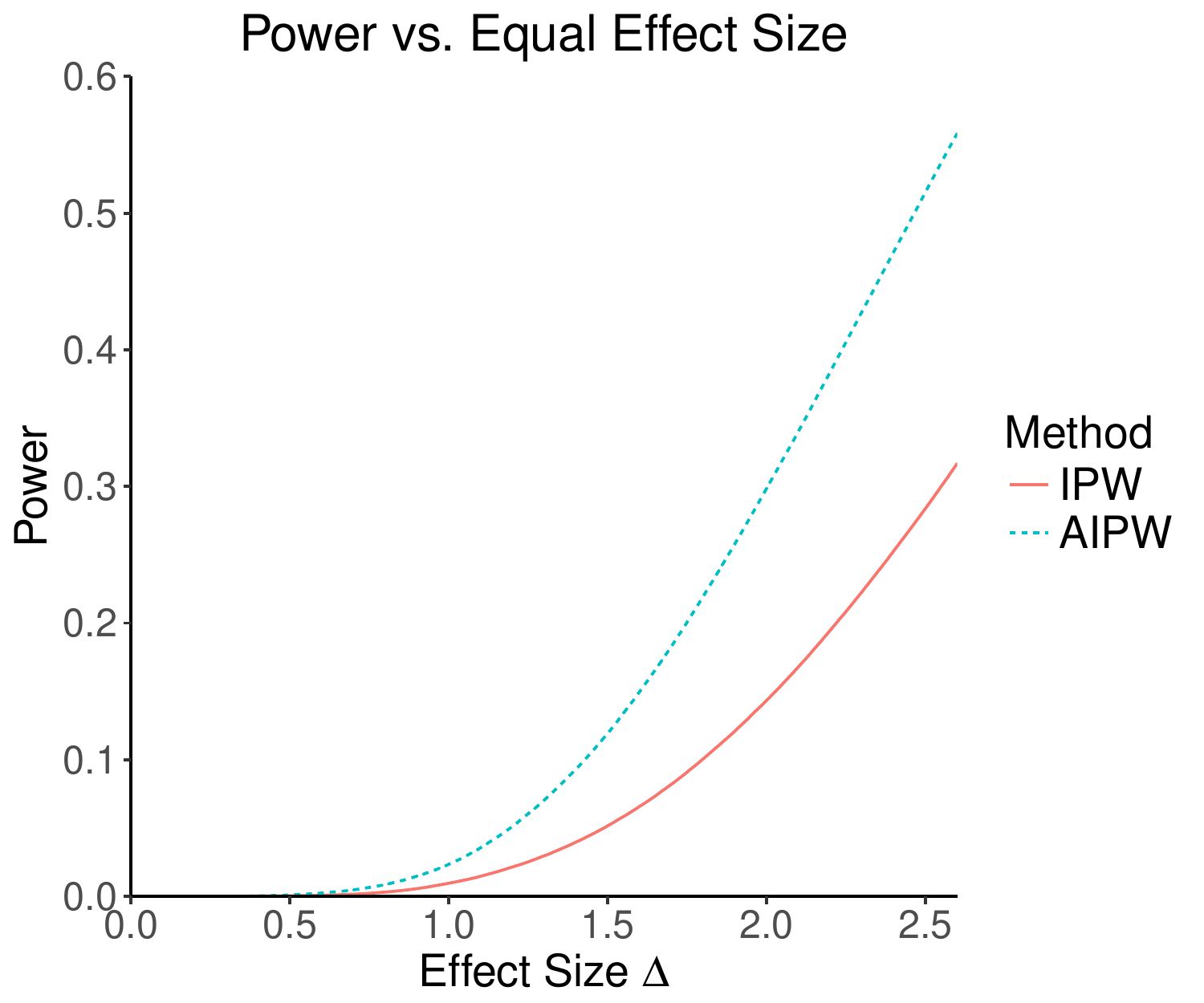}
\caption{This plot shows the power as a function of the uniform effect size in the EXTEND trial when performing estimation with IPW and AIPW, respectively. There are 250 individuals in EXTEND}
\label{fig:EXTEND-power-equal}
\end{figure}

In Figure \ref{fig:EXTEND-power-equal}, we show how the power changes as a function of a uniform effect size. Specifically, we assume $\mathrm{EDTR}_1$ is the best and set the effect sizes of $\mathrm{EDTR}_2,..., \mathrm{EDTR}_8$ to be equal. We then vary this uniform effect size. In this case, we ignore the actual effect sizes of the true EDTR estimates $\hat{\btheta}$. In both Figures \ref{fig:EXTEND-power-step} and \ref{fig:EXTEND-power-equal}, we see the trend that AIPW yields greater power when compared with IPW.

\section{Discussion}
One of the main goals of SMART designs is determination of an optimal dynamic treatment regimen. When planning SMART designs, it is hence crucial to enroll a sufficient number of individuals in order to be able to detect the best DTR and to be able to exclude DTRs inferior to the best one by a clinically significant quantity. 

In this paper, we introduced a novel method for carrying out power analyses for SMART designs which leverages multiple comparison with the best and Monte Carlo simulation. Our methodology directly addresses the central goal of SMARTs.  

We explored the sensitivity of our method to varying parameters in the covariance matrix. We saw that the power prediction is greatly affected by the particular choice of the covariance matrix. This underscores the importance of relying on previous data such as a pilot SMART to estimate $\bSigma$. We saw in simulation studies that our method yields valid estimates of power and it appears similar results may be achieved by choosing exchangeable ``close'' covariance matrices to the true covariance matrix. 

We illustrated our method on the EXTEND SMART to see how much power there was to exclude inferior DTRs from the set of best and the necessary sample size to achieve $80\%$ power.

Future work will involve developing ways of choosing $\bSigma$ using pilot SMART data and for sizing pilot SMARTs with the ability to estimate the covariance matrix to a specified accuracy.

%\section*{Supplementary Material}
%Supplementary Material is available at \url{http://biostatistics.oxfordjournals.org}.

\section*{Acknowledgments}
Conflict of Interest: None declared.

%\section*{Funding}

\section{Figures}

\clearpage

\begin{table}[H]
\centering
\caption{Extend trial: parameter estimates and standard errors}
\begin{tabular}[H]{c|c|cccccccccc}
\hline
 & Parameter& $\theta_1$ & $\theta_2$ & $\theta_3$ & $\theta_4$ & $\theta_5$ & $\theta_6$ & $\theta_7$ & $\theta_8$\\
\hline
IPW & Estimate & 7.56 & 9.53 & 8.05 & 10.02 &  7.71 & 9.68 & 8.19 & 10.17\\
&SD & 0.76 & 0.81 & 0.71 & 0.83 & 0.74 & 0.80 & 0.69 & 0.82 \\
\hline
AIPW & Estimate & 7.65 & 9.44 & 7.83 & 9.62 & 8.06 & 9.85 & 8.24 & 10.03\\
&SD & 0.67 & 0.76 & 0.70 & 0.70 & 0.67 & 0.77 & 0.70 & 0.72\\
\hline
\end{tabular}
\label{tab:1}
\end{table}

\newpage
\section*{Supplementary Material}
\section*{Appendix A.}
\section{Notation}
We focus on notation for two-stage SMART designs, but the methods in this paper are applicable to an arbitrary SMART. Let $\mathrm{EDTR}_i$ denote the $i$th EDTR. Let $O_j$ and $A_j$ denote the observed covariates and treatment assignment, respectively, at stage $j$. Let $\bar{O}_j$ and $\bar{A}_j$ denote the covariate and treatment histories up to and including stage $j$, respectively. Let $\mathcal{T}$ the \textit{treatment trajectory} be the vector of counterfactual treatment assignments for an individual. For example, in a two-stage SMART with response as a tailoring variable, $\mathcal{T}$ may be of the form $\mathcal{T}=(A_1,A_2^{\mathrm{R}}, A_2^{\mathrm{NR}})$ where $A_2^{\mathrm{R}}$ is the stage two treatment assignment had the individual responded and $A_2^{\mathrm{NR}}$ is the stage two treatment assignment had the individual not responded. The reason these are counterfactual treatment assignments is that for an individual who responds to the stage 1 treatment, $A_2^{\mathrm{NR}}$ would be unobserved. Hence, the treatment history $\bar{A}_2$ would be $(A_1,A_2)$ while the treatment trajectory $\mathcal{T}$ would be $(A_1,A_2^{\mathrm{R}}, A_2^{\mathrm{NR}})$ and would include the unobserved counterfactual. Let $Y$ denote the observed outcome of an individual at the end of the study. A tailoring variable may be written as $V$ and is a function of the observed covariates and treatment assignments at each stage. Let $S$ be an indicator for randomization at stage 2. Then, the data structure for a two-stage SMART may be written $(O_1,A_1,O_2,S,A_2,Y)$ (\citealp{ertefaie2015}).
\section{Estimation}
We summarize the estimation procedures IPW (inverse probability weighting) and AIPW (augmented inverse probability weighting) introduced in \cite{ertefaie2015}.
In order to perform estimation with IPW/AIPW, a marginal structural model (MSM) must be specified. An MSM models the response as a function of the counterfactual random treatment assignments captured in the treatment trajectory vector $\mathcal{T}$, while ignoring non treatment covariates. For example, in a two-stage SMART, the MSM is: $$m(\mathcal{T};\boldsymbol\beta) = \beta_0+\beta_1 A_1 +\beta_2 A_2^{R} +\beta_3 A_2^{NR}+\beta_4 A_1 A_2^{R}+\beta_5 A_1 A_2^{NR}.$$
Subsequently, the IPW and AIPW estimators $\hat{\btheta}_{\mathrm{IPW}}$,  $\hat{\btheta}_{\mathrm{AIPW}}$ may be obtained by solving their respective estimating equations: \begin{equation}
\mathbb{P}_n\sum_{k=1}^K\dot{m}(\mathcal{T};\beta)w_2(V,\bar{A}_2,k)(Y-m(\mathcal{T};\beta)=0\tag{IPW}
\end{equation}

\begin{equation}\begin{split}
\mathbb{P}_n\sum_{k=1}^K & \dot{m}(\mathcal{T};\boldsymbol\beta)\big[w_2(V,A_2,k)(y-m(\mathcal{T},\boldsymbol\beta))\\&-\left(w_2(V,\bar{A}_2,k)-w_1(A_1,k)\right)\left(\mathbb{E}[Y \mid \bar{A}_2=\mathrm{EDTR}_k^V,\bar{O}_2]-m(\mathcal{T};\boldsymbol\beta)\right)\\&-(w_1(A_1,k)-1)(\mathbb{E}[\mathbb{E}[Y \mid \bar{A}_2=\mathrm{EDTR}_k^V,\bar{O}_2]\mid A_1 = \mathrm{EDTR}_{k,1},O_1]-m(\mathcal{T};\boldsymbol\beta)\big) = 0
\end{split}\tag{AIPW}\end{equation}
where $\mathbb{P}_n$ denotes the empirical average, 
$\dot{m}(\mathcal{T},\boldsymbol \beta) = \dfrac{\partial m}{\partial\boldsymbol \beta}$, $\mathrm{EDTR}_k^V = \left(\mathrm{EDTR}_{k,1},\mathrm{EDTR}_{k,2}^V\right)$, $w_1(a_1,k) = \dfrac{I_{\mathrm{EDTR}_{k,1}}(a_1)}{p(A_1 = a_1)}$ for $A_1 = a_1$, and $w_2(v,\bar{a}_2, k)=\dfrac{I_{\mathrm{EDTR}_{k,1}}(a_1)I_{\mathrm{EDTR}_{k,2}^v}(a_2)}{p(A_1=a_1)p(A_2=a_2 \mid A_1=a_1, V = v)}$ for $V = v$ and $\bar{A}_2 = \bar{a}_2$.

AIPW is doubly-robust in the sense that it will still provide unbiased estimates of the MSM coefficients $\boldsymbol\beta$ when either the conditional means or the treatment assignment probabilities are correctly specified.

The following proposition from \cite{ertefaie2015} is included for the sake of completeness. 
\begin{prop}
Let $\hat{\theta}^{\diamond}=D\hat{\beta}^{\diamond}$, where $D$ is a $K \times p$ matrix with the $k$th row of $D$ being the contrast corresponding to the $k$th EDTR. Then, under standard regulatory assumptions, $\sqrt{n}(\hat{\theta}^{\diamond}-\theta) \to N(0,\boldsymbol\Sigma^{\diamond}=D[\Gamma^{-1}\Lambda^{\diamond}\Gamma'^{-1}]D')$ where $\Gamma = -\mathbb{E}[\sum_{i=1}^K \dot{m}(\mathcal{T};\beta)\dot{m}'(\mathcal{T};\beta)]$ and $\Lambda^{\diamond}=\mathbb{E}[U^{\diamond}U'^{\diamond}]$ with 
\begin{align*}
U^{\mathrm{AIPW}}&=\begin{aligned}
\sum_{k=1}^K& \dot{m}(\mathcal{T};\boldsymbol\beta)\big[w_2(V,A_2,k)(y-m(\mathcal{T},\boldsymbol\beta))-\left(w_2(V,\bar{A}_2,k)-w_1(A_1,k)\right)\left(\mathbb{E}[Y \mid \bar{A}_2=\mathrm{EDTR}_k^V,\bar{O}_2]-m(\mathcal{T};\boldsymbol\beta)\right)\\&-(w_1(A_1,k)-1)(\mathbb{E}[\mathbb{E}[Y \mid \bar{A}_2=\mathrm{EDTR}_k^V,\bar{O}_2]\mid A_1 = \mathrm{EDTR}_{k,1},O_1]-m(\mathcal{T};\boldsymbol\beta)\big)\end{aligned}\\
U^{\mathrm{IPW}}&=\sum_{k=1}^K\dot{m}(\mathcal{T};\beta)w_2(V,\bar{A}_2,k)(Y-m(\mathcal{T};\beta).
\end{align*}\\
The asymptotic variance $\boldsymbol\Sigma^{\diamond}$ may be estimated consistently by replacing the expectations with expectations with respect to the empirical measure and $(\beta,\gamma)$ with its estimate $(\hat{\beta}^{\diamond},\hat{\gamma})$ and may be denoted as $\hat{\boldsymbol\Sigma}^{\diamond}=D[\hat{\Gamma}^{-1}\hat{\Lambda}^{\diamond}\hat{\Gamma}'^{-1}]D'$
\end{prop}

\section*{Appendix B.}

\section{Proof of Theorem 4.1}
 Note that $\mathrm{Cov}\left(\dfrac{\sqrt{n}(\hat{\theta}_i-\hat{\theta}_N)}{\sigma_{iN}},\dfrac{\sqrt{n}(\hat{\theta}_j-\hat{\theta}_N}{\sigma_{jN}}\right)=\dfrac{\Sigma_{ij}-\Sigma_{iN}-\Sigma_{jN}+\Sigma_{NN}}{\sigma_{iN}\sigma_{jN}}$.

Assume $\bSigma$ is exchangeable, e.g., $\bSigma = \sigma^2\boldsymbol I_{N} + \rho\sigma^2 \left( \mathbbm{1}_N \mathbbm{1}_{N}'-\boldsymbol I_N \right)$ where $\mathbbm{1}_N$ is a vector of $N$ $1's$ and $I_N$ is the $N$ by $N$ identity matrix.

 Then, for all $i,j$, $\mathrm{Cov}\left(W_i,W_j\right)=\mathrm{Cov}\left(\dfrac{\sqrt{n}(\hat{\theta}_i-\hat{\theta}_N)}{\sigma_{iN}},\dfrac{\sqrt{n}(\hat{\theta}_j-\hat{\theta}_N}{\sigma_{jN}}\right) = \dfrac{\rho\sigma^2 -2\rho \sigma^2 +\sigma^2}{2\sigma^2(1-\rho)}=\dfrac{\sigma^2(1-\rho)}{2\sigma^2 (1-\rho)}=\dfrac{1}{2}$ for all $\rho \in \left(-\dfrac{1}{N-1},1\right)$ and for all $\sigma^2$. Also, $c_{i,\alpha}$ is constant across all values of $\rho$ and $\sigma^2$.

 It follows from Slepian's inequality and monotonicity of the probability measure that \\
 $\displaystyle\Pow = \mathbb{P}_{\bSigma}\left(\bigcap_{i=1}^{N-1}\left\{W_i <-c_{i,\alpha}+\dfrac{\Delta_i\sqrt{n}}{\sqrt{2\sigma^2(1-\rho)}}\right\}\right)$

is monotone increasing in $\rho$ and monotone decreasing in $\sigma^2$.
 \\
\section{Simulation Study}
In this section of the supplementary material, we give additional details on how estimation was performed in the simulation study. The SMART designs in the simulation studies are based off the SMART designs of  the simulation studies of \cite{ertefaie2015}. We estimated $\btheta$ and $\bSigma$ using AIPW\\
For SMART design 1, the MSM is $m(\mathcal{T};\beta)=\beta_0+\beta_1A_1+\beta_2A_2^{\mathrm{NR}}.$
The true conditional means are:
\begin{align*}
\mathbb{E}[Y \mid \bar{A}_2=\mathrm{EDTR}_k^V,\bar{O}_2] &= \gamma_0 +\gamma_1o_{11}+\gamma_2o_{12}+\gamma_3o_{21}+\gamma_4o_{22} + a_1(\gamma_5+\gamma_6o_{11})+\gamma_7I(O_{21}>0))a_2\\
\mathbb{E}[Y\mid A_1 = \mathrm{EDTR}_{k,1},O_1] &= \gamma_8+\gamma_9o_{11}+\gamma_{10}o_{12}+\gamma_{11}a_1+\gamma_{12}a_1o_{11}
\end{align*}

For SMART design 2, the MSM is: 
\begin{equation*}
m(\mathcal{T};\boldsymbol\beta) = \beta_0+\beta_1I(A_1=-1)+I(A_1 = -1)[\beta_2I(A_2^B = 1)+\beta_3I(A_2^B=2)+\beta_4I(A_2^B=3)].
\end{equation*} 

The true conditional means are: 
\begin{align*}
\mathbb{E}[Y \mid \bar{a}_2=\mathrm{EDTR}_k^V,\bar{o}_2,\boldsymbol\gamma] = &\gamma_0+\gamma_1o_{11}+\gamma_2o_{12}+\gamma_3o_{21}+\gamma_4o_{22}+I(a_1=-1)(\gamma_5+\gamma_6o_{11}) \\
&+I(o_{21}>0)I(a_1=-1)[\gamma_7I(a_2=1)+\gamma_8I(a_2=2)+\gamma_9I(a_2=3)+\gamma_{10}o_{21}I(a_2=2)]\\
\mathbb{E}[Y\mid a_1 \in \mathrm{EDTR}_{k,1},o_1,\boldsymbol\gamma] = &\gamma_{11}+\gamma_{12}o_{11}+\gamma_{13}o_{12}+\gamma_{14}I(a_1=-1)+\gamma_{15}I(a_1=-1)o_{11}
\end{align*}

\begin{figure}[t]
\centering
\includegraphics[width = 15cm]{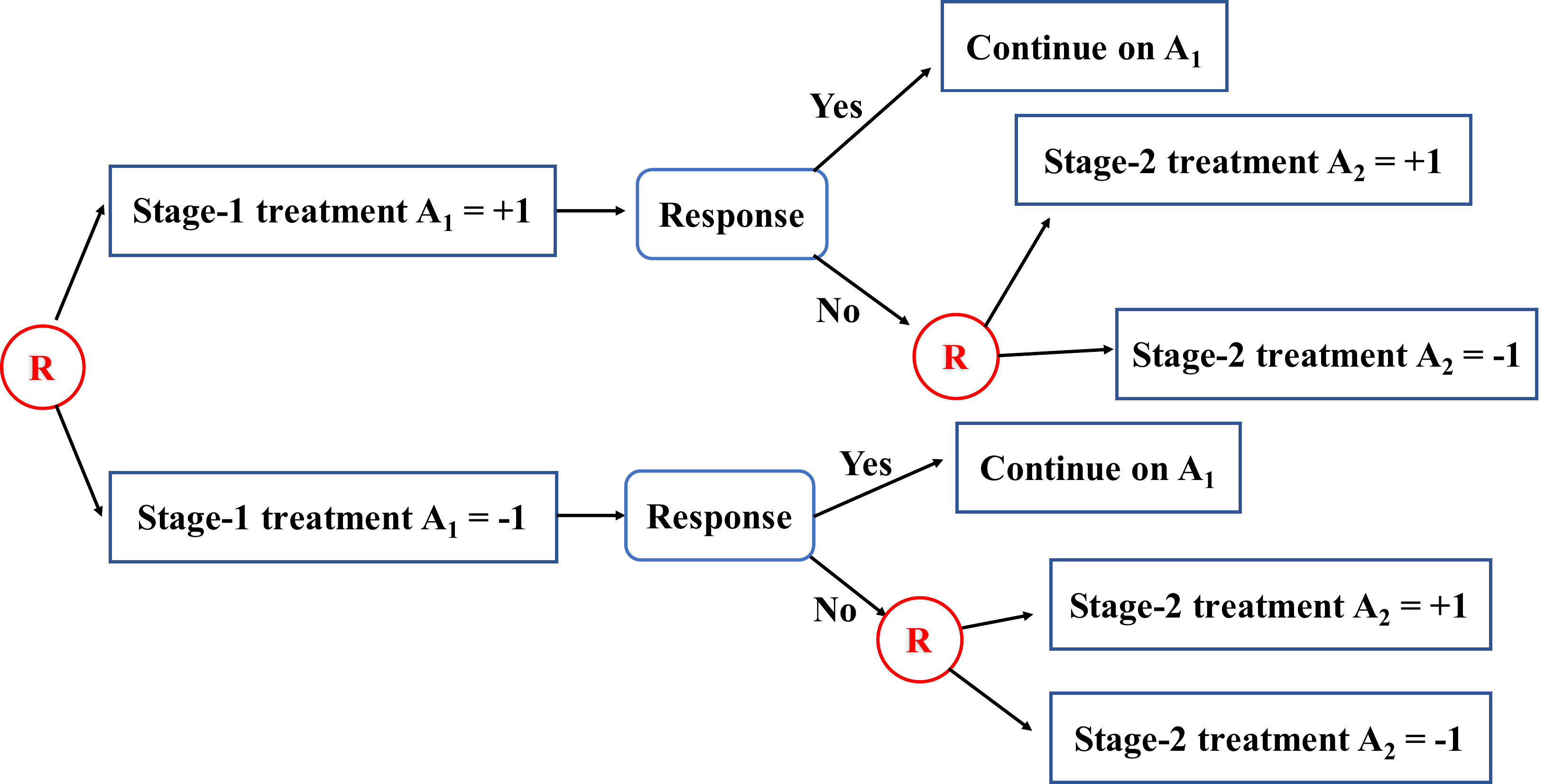}
\caption{SMART Design 1}
\label{fig:SMART-design-1}
\end{figure}

\begin{figure}[t]
\centering
\includegraphics[width = 15cm]{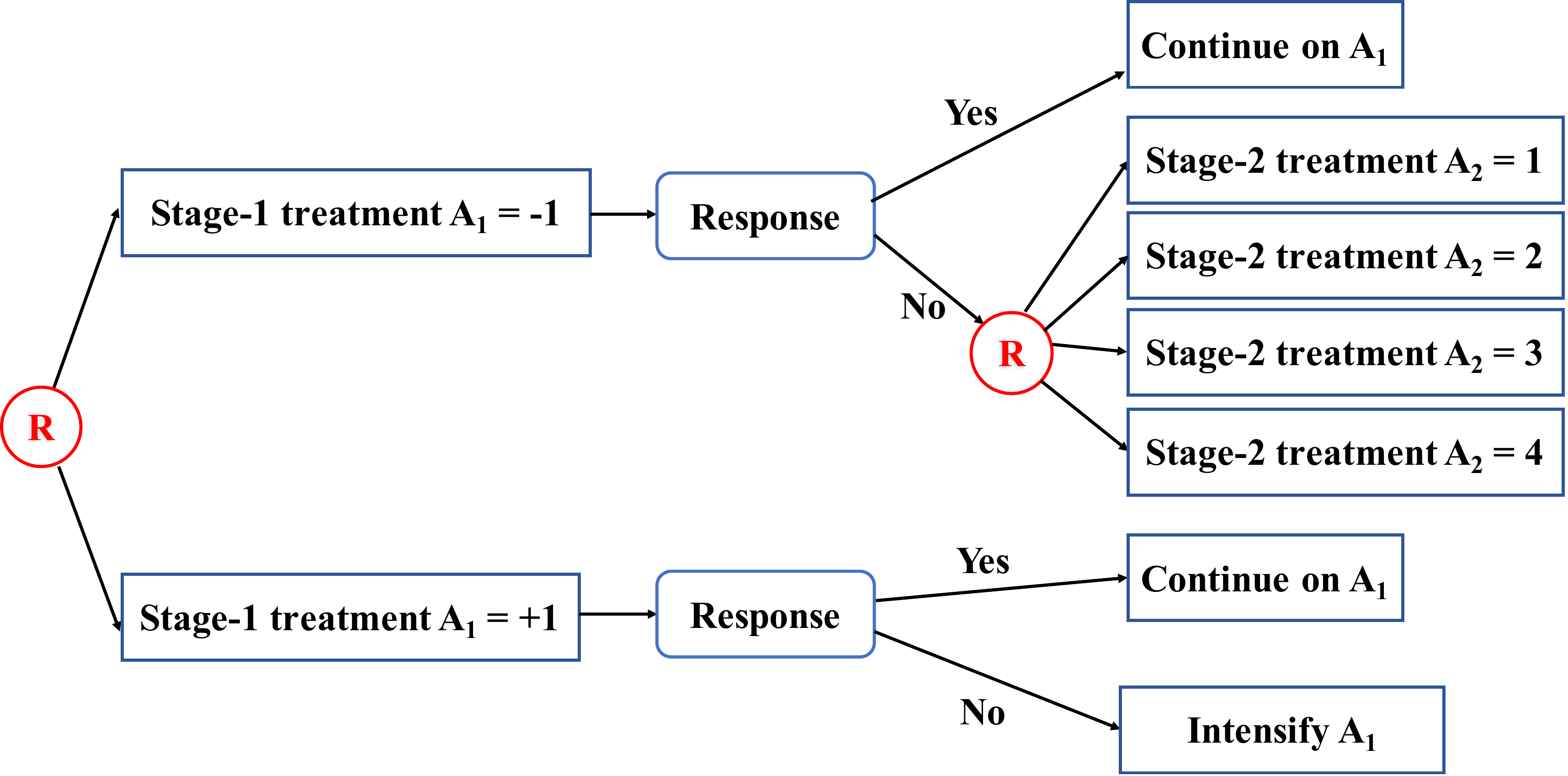}
\caption{SMART Design 2}
\label{fig:SMART-design-2}
\end{figure}

\subsection{Simulation SMART design 1: Determination of the Closest Exchangeable Matrix}
Let $\boldsymbol\Sigma$ be exchangeable and $\lVert\cdot\rVert$ denote the Frobenius norm. Then, $\boldsymbol\Sigma$ is of the form $\boldsymbol\Sigma = \sigma^2\boldsymbol I_{N} + \rho\sigma^2 \left( \mathbbm{1}_N \mathbbm{1}_{N}'-\boldsymbol I_N \right)$. Hence, 
$$\left \lVert \boldsymbol\Sigma-\boldsymbol\Sigma_{\mathrm{True}}\right\rVert^2 = \left\lVert \sigma^2\boldsymbol I_{N} + \rho\sigma^2 \left( \mathbbm{1}_N \mathbbm{1}_{N}'-\boldsymbol I_N \right) - \boldsymbol\Sigma_{\mathrm{True}}\right\rVert^2 := \sum_{i = 1}^N (\sigma^2-\sigma_i^2)^2 + \underset{i\neq j}{\sum\sum}(\rho\sigma^2 - \sigma_{ij})^2$$
Then, \begin{align*} \dfrac{\partial}{\partial \sigma^2}\left\lVert \boldsymbol\Sigma-\boldsymbol\Sigma_{\mathrm{True}}\right\rVert^2 &= 2\sum_{i=1}^N(\sigma^2-\sigma^2_i)+2\rho\underset{i\neq j}{\sum\sum}(\rho\sigma^2-\sigma_{ij})=0 \\ \text{and }\dfrac{\partial}{\partial \rho}\left\lVert \boldsymbol\Sigma-\boldsymbol\Sigma_{\mathrm{True}}\right\rVert^2 &=2\sigma^2 \underset{i\neq j}{\sum\sum}(\rho\sigma^2-\sigma_{ij})=0. \end{align*}
Hence, 
$$\sigma^2 = \dfrac{1}{N}\sum_{i=1}^N\sigma_i^2 \text{ and }  \rho\sigma^2 = \dfrac{1}{N(N-1)}\underset{i \neq j}{\sum\sum} \sigma_{ij}$$ 

\subsection{Simulation 2: Determination of the closest exchangeable matrix}
We assumed a block exchangeable matrix for $\bSigma$ of the form \begin{equation*}\boldsymbol\Sigma_{\mathrm{Exchangeable}}=\begin{pmatrix}\sigma_{1w}^2 &\rho_1\sigma_{1w}\sigma_{2w}&\rho_1\sigma_{1w}\sigma_{2w}&\rho_1\sigma_{12}\sigma_{2w}&\rho_1\sigma_{1w}\sigma_{2w}\\
 \rho_1\sigma_{1w}\sigma_{2w} &\sigma_{2w}^2&\rho_2\sigma_{2w}^2&\rho_2\sigma_{2w}^2&\rho_2\sigma_{2w}^2\\
 \rho_1\sigma_{1w}\sigma_{2w}&\rho_2\sigma_{2w}^2&\sigma_{2w}^2&\rho_2\sigma_{2w}^2&\rho_2\sigma_{2w}^2\\
 \rho_1\sigma_{1w}\sigma_{2w}&\rho_2\sigma_{2w}^2&\rho_2\sigma_{2w}^2&\sigma_{2w}^2&\rho_2\sigma_{2w}^2\\
 \rho_1\sigma_{1w}\sigma_{2w}&\rho_2\sigma_{2w}^2&\rho_2\sigma_{2w}^2&\rho_2\sigma_{2w}^2&\sigma_{2w}^2
 \end{pmatrix}
 \end{equation*}
 Note that
 $$\left\lVert \boldsymbol\Sigma_{\mathrm{Exchangeable}}-\boldsymbol\Sigma_{\mathrm{True}}\right\rVert=(\sigma_{1w}-\sigma_1^2)^2+\sum_{i=2}^5(\sigma_{2w}^2-\sigma_2^2)^2 +2\sum_{j=2}^5(\rho_1\sigma_{1w}\sigma_{2w}-\sigma_{1j})^2+\underset{i\neq j \in \{2,3,4,5\}}{\sum\sum}(\rho_2\sigma_{2w}^2-\sigma_{ij}).$$
 
 Hence, 
 
 \begin{align*}
 \dfrac{\partial}{\partial\sigma_{1w}^2} \lVert\boldsymbol\Sigma-\boldsymbol\Sigma_{\mathrm{close}}\rVert &= 2(\sigma_{1w}^2-\sigma_1^2) + 4\rho_1\sigma_{2w}\sum_{j=2}^5(\rho_1\sigma_{1w}\sigma_{2w}-\sigma_{1j}) = 0\\
 \dfrac{\partial}{\partial\rho_1}\lVert\boldsymbol\Sigma-\boldsymbol\Sigma_{\mathrm{close}}\rVert &=4\sigma_{1w}\sigma_{2w}\sum_{j=2}^5(\rho_1\sigma_{1w}\sigma_{2w}-\sigma_{1j}) = 0\\
 \dfrac{\partial}{\partial\sigma_{2w}^2}\lVert\boldsymbol\Sigma-\boldsymbol\Sigma_{\mathrm{close}}\rVert &=2\sum_{i = 2}^5(\sigma_{2w}^2-\sigma_i^2)+4\rho_1\sigma_{1w}\sum_{j=2}^5(\rho_1\sigma_{1w}\sigma_{2w}-\sigma_{1j})+2\rho_2\underset{i\neq j, 1}{\sum\sum}(\rho_2\sigma_{2w}^2-\sigma_{ij})\\
 \dfrac{\partial}{\partial\rho_2}\lVert\boldsymbol\Sigma-\boldsymbol\Sigma_{\mathrm{close}}\rVert &= 2\sigma_{2w}^2 \underset{i\neq j,1}{\sum\sum}(\rho_2\sigma_{2w}^2-\sigma_{ij})
 \end{align*}
 Hence, 
 \begin{align*}
 \sigma_{1w}^2 &= \sigma_{1w}\\
 \sigma_{2w}^2 &= \dfrac{1}{4}\sum_{j=2}^5 \sigma_j^2\\
 \rho_1\sigma_{1w}\sigma_{2w}&= \dfrac{1}{4}\sum_{j=2}^5\sigma_{1j}\\
 \rho_2\sigma_{2w}^2 &= \dfrac{1}{12}\underset{j\neq i \in \{2,3,4,5\} }{\sum\sum}\sigma_{ij}
 \end{align*}
\clearpage
\bibliographystyle{biorefs}
\bibliography{refs}
\clearpage
\end{document}